\documentclass[aps,nofootinbib,showpacs,preprintnumbers]{revtex4}
\usepackage{epsfig}

\begin{document}
\preprint{hep-ph/0207291, v3, Oct.~2002, to appear in PRD}

\title{Resummations of free energy at high temperature}

\author{G.~Cveti\v{c}}
  \email{gorazd.cvetic@fis.utfsm.cl}
\affiliation{Department of Physics, Universidad T\'ecnica
Federico Santa Mar\'{\i}a, Valpara\'{\i}so, Chile}
\author{R.~K\"ogerler}
 \email{koeg@physik.uni-bielefeld.de}
\affiliation{Department of Physics, Universit\"at Bielefeld, 
33501 Bielefeld, Germany}

\date{\today}

\begin{abstract}
We discuss resummation strategies for free energy in 
quantum field theories at nonzero temperatures $T$. 
We point out that resummations should be performed 
for the short- and long-distance parts separately in order to avoid
spurious interference effects and double-counting.
We then discuss and perform Pad\'e resummations of these two 
parts for QCD at high $T$. The resummed results are
almost invariant under variation of the
renormalization and factorization scales. 
We perform the analysis also in the case
of the massless scalar $\phi^4$ theory.
\end{abstract}
\pacs{12.38.Cy, 11.10.Wx, 12.38.Bx, 12.38.Mh}

\maketitle

\section{Introduction}
\label{sec:introduction}

During the last decade the physics of many-particle
systems at relativistic energies has attracted
increasing interest. On the one hand, this was due to
the hope that high-energy ion collisions at
present day colliders would reach the kinematical
regime of the quark-gluon-plasma phase of QCD and
would thus give new insight into the phase structure 
of QCD and into the confinement phenomena.
On the other hand, a better understanding of such systems
can be applied to ultrarelativistic astrophysical and/or
cosmological configurations and should therefore provide a
conceptual tool for describing the physics of the early
universe, in particular phenomena such as baryon number
generation, big-bang nucleosynthesis, etc. Although it
is not yet clear to what extent the various mentioned
configurations can be considered as being in thermal
equilibrium, the first attempts at a systematic analysis
concentrated on equilibrium statistical mechanics.

Since we are convinced that the interaction of particles
at energies up to 1 TeV is described by the standard
model of strong and electroweak interactions, the main
task consists in formulating a consistent thermal quantum
field theory of QCD and QED. Because QCD is asymptotically
free, the corresponding running coupling parameter 
becomes small at a sufficiently high temperature $T$
($\alpha_s(T) \ll 1$)\footnote{
$T$ being considered as a measure of the average energy
of the system's constituents.}
and a perturbative treatment seems possible.
This expectation is supported by the observation that
lattice gauge theory calculations 
\cite{Boyd:1995zg,Engelsetal96,Karsch:2000ps}
indicate that the physics of the (quark-)gluon plasma
at $T$ larger than four times the critical temperature
$T_c$ is to a very good approximation that of an ideal
gas. In particular, the free energy $F$ of such a system 
is off the ideal gas value by less than $20 \%$.
Even more, if masses with an appropriate 
temperature dependence are accepted, 
the lattice results \cite{PeshierLevai}
agree almost exactly with the ideal gas ones for
all values of $T > T_c$. 
This would suggest that perturbative treatments
remain candidates for a description of the relevant phenomena.

Unfortunately, a straightforward perturbative treatment
has led to results which are far from satisfactory.
The problems emerging can be seen best in the expressions
for the free energy $F$ as calculated in various theoretical
models. The most exhaustive calculations, up to $\sim\!g^5$,
have been performed for the massless $\phi^4$ theory
\cite{FrenkelParwani,Braaten:1995cm}
and for QCD (without and with quarks)
\cite{ArnoldZhai,Braaten:1996ju}.
These results have the following unpleasant features:
\begin{enumerate}
\item
There is a very bad convergence behavior of the series,\footnote{
Formally, the series is believed to be an 
asymptotically divergent series.} 
some of the higher-order terms in the perturbation series
are larger than the leading term unless the 
coupling parameter is very small -- in QCD,
the coupling $\alpha_s(T)$ would have to be
at $T > 1$ TeV. If successive orders are included 
in the truncated perturbation series (TPS), 
the sum takes on alternating large values
(cf. Fig.~1 of Ref.~\cite{Andersen:1999sf}).
\item
At every fixed order the TPS's show a strong
dependence of the (arbitrary) renormalization
scale (RScl) $\mu$. This results in an additional
severe uncertainty of the evaluation, this time
due to the ambiguity of the choice of the RScl $\mu$.
In an asymptotically free theory one usually is 
on the safe side if one chooses $\mu$ close to
the (lowest) physically relevant momentum scale
typical of the contributing configurations,
since otherwise the expansion coefficients would
blow up. In the case of a system at temperature $T$
one would be led to choose $\mu\!\sim\!T$, or
more specifically $\mu\!=\!2 \pi T$, i.e., the energy
of the first nonvanishing Matsubara mode. In fact,
this is the choice actually used in the literature.
However, we should keep in mind that in the
case of a plasma consisting of massless components
there are important physical effects connected with much
lower energies (screening of the chromoelectric or
chromomagnetic field). Therefore, the choice $\mu \approx
2 \pi T$ seems not to be well founded, {\em a priori}.
\end{enumerate}

These features are related to the rather large genuine
collective effects. And, in fact, it has been noted
recently that these unexpectedly large corrections
are rooted in the terms $\sim\!g^3, g^5$
within the expansions. Such terms render the expression
nonanalytic in $a\!\equiv\!g^2/(4 \pi^2)$ and they do not 
emerge in the ordinary ($T\!=\!0$) field theory.
They are the result of a certain (partial) resummation
which is necessary to get rid of these infrared divergences 
\cite{Kapusta,Karsch:1997gj}.
They are closely related with collective effects such as
screening or Landau damping.

This observation has constituted the basis for several
attempts to remedy the situation: since a specific method of
resummation seems to be the root of the problems, one
is led to try different resummation procedures.
The main attempt goes under the name of
screened perturbation theory (SPT)
\cite{Karsch:1997gj,Andersen:2000yj}
and has been made more systematic by what is called
optimized perturbation theory
\cite{Chiku:1998kd}. The main idea is to 
add a local mass term to the free part
of the Lagrangian and subtract it from
the interaction part. The former is treated
nonperturbatively, constituting a genuine 
(screening) mass; the latter perturbatively.
Physically this means that one expands about a
(ideal) gas of massive quasiparticles (dressed
gluons, for example) rather than about massless
particles. The technical consequences are
striking. Since the objects one starts with
are massive, there is no infrared problem and
no need for resummation, and the resulting
expressions have better convergence behavior than
the original perturbation series. However,
for obtaining numerical results, one has to fix the
chosen mass at the final stage. 
The authors of Ref.~\cite{Karsch:1997gj}, e.g., used 
(an appropriate approximation of) the gap equation
for fixing this mass. 

All this is relatively
straightforward in a theory such as $\phi^4$ where
nothing forbids a genuine mass term. 
Within a gauge theory like QCD the whole procedure becomes
much more cumbersome, since the addition of a genuine
(local) mass term would spoil the gauge symmetry
from the outset. A technical way out of this
problem is hard-thermal loop (HTL) perturbation theory
\cite{Andersen:1999sf}. This is a SPT generalization
which respects gauge invariance. The procedure again rests 
on a (partial) resummation: Those higher-order loop corrections
are included that are of leading order in $g_s$ for 
amplitudes involving soft external momenta $p\sim g_s T$.
Unfortunately, the resulting HTL correction terms are 
nonlocal. This leads to complicated UV divergences,
and only some of them are canceled by physical mechanisms
(quasiparticle formation and Landau damping). The rest
have to be tamed by artificial counterterms proportional
to the quasiparticle mass and they generate an additional
renormalization scale dependence. Despite this complicated
and not uniquely specified procedure, the resulting
series has again a better convergence behavior than
the original perturbation series. 
On the other hand, when the so called
$\Phi$-derivable approximation scheme 
\cite{BIR,Peshier:2000hx} is applied,
some of the problems of the HTL perturbation
theory are avoided. 
In this scheme, the HTL contributions to the
free energy were resummed in Ref.~\cite{Peshier:2000hx} 
in $\phi^4$ theory, QED, and QCD,
but the UV divergences were shown to cancel.
A related approach which uses the framework of
the scalar $O(N)$-symmetric model in the large-$N$
limit was developed in Ref.~\cite{Drummond:1997cw}. 
Another approach, overcoming the infrared problems
appearing in the conventional perturbation theory
and using dressed propagators,
was developed in Ref.~\cite{Drummond:1999si}. 

On the other hand, the strong renormalization scale (RScl)
dependence is not or not sufficiently reduced by
the SPT or HTL approach. Further, there remains a
principal question: Why are the SPT results so much
better than those of the naive perturbation approach?
Although the need for resummation is avoided in the
first step in SPT due to the mass $m_g$ of the
quasiparticle, the correct choice of the specific
formula for $m_g$ again needs a resummation implicitly
contained in the gap equation. As long as these points are
not clarified, the SPT treatment cannot be considered
fully satisfying.

A completely different way of remedying the weak
points of the naive perturbation theory consists
in replacing the truncated perturbation series (TPS)
by appropriate Pad\'e approximants (PA's). This approach
is motivated by at least two features of PA's: 
First, PA's at increasing order 
in general show much better convergence than the TPS's
from which they are obtained. It has been shown that
even when the TPS's are divergent (asymptotic series), the
corresponding PA's may converge and do it under
rather general conditions \cite{Padebook}.
Secondly, PA's reduce considerably the RScl dependence
of the TPS. In fact, it is known that the diagonal PA's,
constructed from a TPS of an RScl-independent
quantity in powers of $a(\mu)\!\equiv\!g^2(\mu)/(4 \pi^2)$,
are RScl-independent in the limit of the one-loop
running of the coupling parameter $a(\mu)$
(large $\beta_0$-limit) \cite{Gardi:1996iq}.
Further, related approximants 
have been developed which are exactly RScl-independent
\cite{CK2} and even renormalization scheme
independent \cite{CK1}. Since the full
perturbation series corresponds to a physical
(in principle measurable) quantity, this
quantity is exactly RScl-independent.
Therefore, one is led to conjecture that
PA's and related approximants are nearer to the true 
value than the original TPS's. The physical
reason for the (approximate) RScl independence
of such approximants is that they include a certain
resummation \cite{Gardi:1996iq}, thus containing infinitely many terms
whose absence was responsible for the
spurious RScl-dependence of the TPS.

The first applications of PA's to thermal
perturbation theory were made by Kastening 
\cite{Kastening:1997rg}
and by Hatsuda \cite{Hatsuda:1997wf}. 
These authors started from the the available
TPS for the free energy $F$ (in $\phi^4$ and QCD),
which is a TPS in powers of $a^{1/2}\!=\!g/(2 \pi)$,
up to $\sim\!a^{5/2}$, and replaced it by
various PA's based on it. They demonstrated
improved RScl stability of these PA's.
Although the relatively large number of powers in
$g$ of the underlying TPS suggests a good
convergence quality of the resulting PA's,
the results have to be treated with caution.
First, there is no formal reason to expect
that certain PA's, specifically the diagonal ones,
should be stable under variation of the RScl.
In this respect, we note that it is the one-loop
(large-$\beta_0$) running of $a\!=\!g^2/(4 \pi^2)$,
\begin{equation}
a(\mu_1) = \frac{a(\mu_0)}{1 + a(\mu_0) \beta_0 
\ln(\mu_1^2/\mu_0^2)} \ ,
\label{1lrun}
\end{equation}
which is responsible for the (large-$\beta_0$) 
RScl independence in diagonal PA's of underlying TPS's 
in powers of $a$. This is so because relation
(\ref{1lrun}) represents a homographic transformation
$a \mapsto a/(1 + K a)$ \cite{Gardi:1996iq}. 
However, relation (\ref{1lrun}) yields a structurally
different one-loop running of $g\!=\!2 \pi a^{1/2}$, namely
\begin{equation}
g(\mu_1) = \frac{g(\mu_0)}{\left[ 1 + g^2(\mu_0) \beta_0^{\prime}
\ln(\mu_1^2/\mu_0^2) \right]^{1/2} } \ , \qquad 
(\beta_0^{\prime} \equiv \frac{\beta_0}{4 \pi^2}) \ .
\label{1lgrun}
\end{equation}
This relation is not homographic, and consequently
the PA's in such a case need not be (large-$\beta_0$)
RScl-invariant.

There is a second, more significant point of
criticism of the aforementioned PA approach.
It can be best explained in terms of diagrams.
As we have mentioned, the PA's represent a
resummation (analytic continuation) of the infinite 
sum of a certain class of diagrams. On the other hand,
the TPS expressions for $F(T)$ also include a selective
resummation (of ring and super-ring diagrams).
Therefore, a naive application of PA's to the original
TPS constitutes a mixing of (at least) two inequivalent
resummation effects and could easily lead to partial
double-counting or other (interference-like)
inconsistencies. In the context of QED and QCD
at $T\!=\!0$, somewhat similar aspects have been
pointed out and accounted for in Refs.~\cite{lightbylight}.

Within the present paper we want to stick to PA's
because of their unique advantages: reduced RScl-dependence,
better convergence properties, and (quasi)analytic continuation
aspects. However, we are going to develop a procedure
that is free of the aforementioned weaknesses. We
do this by separating all the terms in the
full available TPS into groups which represent TPS's of
separate physical, i.e., RScl-independent, quantities.
To each of the TPS's we apply PA's separately. The resulting
expressions are not only (approximately) RScl-independent,
but are supposed to represent better approximations to the
true values since double-counting and interference-like
inconsistencies are excluded.

In Sec.~II we present the main elements of our method.
In Sec.~III we apply this method to QCD, and
in Sec.~IV to the massless $\phi^4$ theory. Section V
summarizes the results and presents conclusions.

\section{Separation of long and short distance regimes}
\label{sec:separation}

We start with the perturbatively calculated
expressions for the free energy density $F$
both for massless $\phi^4$ and for QCD.
They have been calculated recently 
\cite{FrenkelParwani,Braaten:1995cm,ArnoldZhai,Braaten:1996ju} 
up to $\sim\!g^5$, i.e.,
three loops plus all the ring diagrams summed up.
The generic structure of the resulting expressions is
\begin{eqnarray}
F(T) &=& F_{\rm ideal} \left[ C_0 + C_2 a + C_3 a^{3/2} +
C_4(\mu) a^2 + C_5(\mu) a^{5/2} + {\cal O}(a^3 \ln a)
\right] \ .
\label{genericF}
\end{eqnarray}
Here, $a$ is short-hand for $a(\mu)\!=\!g^2(\mu)/(4 \pi^2)$,
i.e., the running (RScl-dependent) coupling parameter.
The renormalization scheme (RSch) is assumed to be fixed, say
${\overline {\rm MS}}$. The RScl-running is described
by the perturbatively specified renormalization group
equation (RGE)
\begin{equation}
\frac{ \partial a(\mu^2) }{\partial \ln \mu^2 } = - \beta_0 a^2
- \beta_1 a^3 - \beta_3 a^4 - \ldots \ .
\label{RGE1}
\end{equation}
It is interesting to note that in the series
(\ref{genericF}) the coefficients $C_0, C_2, C_3$
do not depend on the RScl $\mu$, but $C_4$ and $C_5$ do,
the $\mu$-dependence showing up as an additive
contribution $\propto\!\ln[\mu/(4 \pi T)]$ to the
corresponding coefficient. In QCD, $C_4$
includes a term $\propto\!\ln[a(\mu)]$ [$\sim\!1/\ln\ln (\mu)$];
in $\phi^4$ theory, $C_5$ includes a term
$\propto\!\ln[a(\mu)]$.
It is crucial to keep in mind the origin of the
various terms: The terms $\sim\!a^{3/2}$ and
$\sim\!a^{5/2}$ come exclusively from
the resummation of the ring diagrams that is
necessary to avoid infrared divergences
\cite{FrenkelParwani,ArnoldZhai}. 
This resummation procedure also yields part of the
coefficient $C_4$ of the $\sim\!a^2$ term. 
In particular, the contributions $\propto\!\ln(a)$
in $C_4$ and $C_5$ are generated by this procedure
as the remnant of the emerging dependence on
the screening mass. The other terms ($C_0, C_2$, and
the remaining part of $C_4$) stem from ordinary
perturbation theory in powers of $a$ and do not
contain resummation effects. We further note that
the term $\sim\!a^3$ ($\sim\!g^6$) cannot be
obtained by perturbative methods (which include
ring summation) because of the severe
infrared divergences appearing at that order
\cite{Gross:1980br}.

Our aim is to apply Pad\'e approximants (PA's) in a 
consistent manner to a TPS of the type (\ref{genericF}).
As argued in the Introduction, in order to avoid
an uncontrollable mixing and superposition of
different resummations, the separation of the pure 
perturbative from the ring resummation-generated terms
should be performed. In addition, however, we want to 
take advantage of the approximate
$\mu$-independence of the PA's when they are
applied to TPS's (in $a$) of $\mu$-independent
quantities. Therefore, the right-hand side of
Eq.~(\ref{genericF}) should be split
so that the resulting parts (infinite power
series) represent quantities that are $\mu$-independent 
("physical") separately. This suggests
that a physical principle of separation should be 
involved, i.e., one which is connected with
measurable effects. 

In finding such a separation principle
one is guided by the decomposition of all
thermodynamic quantities in (Fourier) modes.
Within the imaginary time formalism, the free
energy $F$ at every given order is expressed
as a sum over the Matsubara frequencies $\omega_n$
\begin{eqnarray}
\omega_n &=& 2 \pi T n  \qquad {\rm for \ bosons} \ ,
\label{Matsubos}
\\
         &=&  \pi T (2 n + 1) \qquad {\rm for \ fermions} \ ,
\label{Matsufer}
\end{eqnarray}
where $n\!=\!0,1,\ldots$.
Since the thermodynamic quantities and correlation
functions can be derived from $F$, they also show up
as sums over modes. For a given correlation
function the contribution from the (exchange of the)
Fourier mode with frequency $\omega_n$ falls off
at large spatial distances as $\exp(- \omega_n R)$
(at least if $T$ is larger than all contributing masses).
Therefore, the only mode which does not produce
an exponentially vanishing contribution to the
long-range correlation functions is the bosonic 
zero mode $\omega_0\!=\!0$ [Eq.~(\ref{Matsubos})].
Consequently, at sufficiently high $T$, the static
correlators of the contributing fields at large
distances $R \gg 1/T$ are exclusively determined by the
zero mode. Since this long-distance behavior
of correlators is, at least in principle,
observable ("physical"), the procedure of separating
the bosonic zero-mode (long-distance) contributions
from all the other (short-distance) ones rests
on physical grounds. We therefore expect that both contributions
are separately $\mu$-independent since both have
a physical meaning. Further, we know that all
resummation effects in the series (\ref{genericF})
contribute exclusively to the long-distance part.
This is so because these resummation effects of the
ring diagrams exclusively stem
from the zero-mode contributions, which represent the
strongest infrared divergences at the single diagram
level. The long-distance part, i.e., the resummed ring
diagrams, shows up as a power series in powers of
$g \propto a^{1/2}$, starting with the $a^{3/2}$ term.
The short-distance part, on the other hand, has
the ordinary perturbation character -- a perturbation
series in integer powers of $a\!\equiv\!g^2/(4 \pi^2)$.

The discussed decomposition of $F$ represents
the basis for our improvement of the
underlying TPS results: We apply the appropriate
PA's to the separate parts, so that (at least some
of) the approximants are approximately RScl-independent,
and presumably better converging. In the case of the
short-range contributions these are the diagonal PA's
$[n/n](a)$. At the available order of the underlying
TPS (\ref{genericF}), the short-range part is of
the form $F_{\rm S}(T) = F_{\rm ideal} ( C_0 + C_2 a +
{\bar C}_4 a^2 )$, and the only possible diagonal
PA is $[1/1](a)$.\footnote{
The Pad\'e approximant $[n/m](x)$ 
for a quantity $S(x)$ is a ratio of two 
polynomials of $n$'th and $m$'th degree in $x$: 
$[n/m](x) = P_n(x)/P_m(x)$, such that its expansion
in powers of $x$ reproduces the terms up to $\sim\!x^{m+n}$
of the expansion of $S(x)$:
$S(x) = r_0\!+\!r_1 x\!+\!\ldots r_{n+m} x^{n+m}\!+\!\ldots$.}  
In the case of the long-range contributions,
one tries to see which PA $[n/m](g)$ ($n+m=2$) is
approximately RScl-independent.

Before continuing, we mention an alternative way
to interprete the described decomposition
on physical grounds: The long-distance contribution
reflects the observable phenomena of screening
of interactions and formation of quasiparticles
(collective modes). 
To understand this, we recall the following facts.
It is with the resummation of the ring diagrams
that the zero modes acquire a (screening) mass
$m_s\!\sim\!g T$ ($m_s$ can be defined as the
solution of the gap equation \cite{Karsch:1997gj,Andersen:2000yj}).
And vice-versa: It is the screening phenomenon
which explains the striking fact that, while all
individual ring diagrams are infrared divergent,
their total sum is convergent. 
{}From this one concludes that,
due to the observable nature of the screening phenomenon,
the long-range part $F_{L}$, which encompasses the
resummation-generated effects, is of physical
nature and should therefore be RScl-independent.
This can best be seen by
determining the screening mass via the gap equation
\cite{Karsch:1997gj,Andersen:2000yj}
which is equivalent to summing up the ring diagrams.

This interpretation also shows that in theories
with more degrees of freedom (like QCD) there
might be several stages of screening, e.g., screening of
chromoelectric and chromomagnetic gluons, respectively.
Consequently, the free energy $F$ can then be
decomposed into more than two parts: the short-distance
part, and one part for each kind of screening.

Having described the physical idea behind our
approach, we have not yet addressed the technical problem 
of actual decomposition. In principle, the answer is
simple: in order to single out the long-range
(zero-mode) contribution, one has to integrate out
all higher modes (bosonic, and all fermionic modes). 
The resulting expression,
when subtracted from the full expression,
should yield the short-distance contribution,
i.e., that of all the nonzero modes. In practice,
integrating out explicitly all higher modes
is a cumbersome task. Fortunately,
there exists an alternative method,
the method of effective field theories,
which was developed for thermal perturbation
theory by Braaten and Nieto
\cite{Braaten:1994na,Braaten:1995cm,Braaten:1996ju}. 
The method can be briefly described as follows.
At small distances ($R \leq 1/T$) the behavior
of the system is determined by ordinary perturbative
QCD. On the other hand, 
the long-distance behavior is dominated by
the zero modes and can, consequently, be described
by an effective field theory which, at large $T$,
is a bosonic field theory in three dimensions
("dimensional reduction" \cite{Gross:1980br,Appelquist}). 
The effective bosonic field
is approximately identified with the zero modes of the
original fields. For the construction of the effective
field theory one does not need to specify this
effective (static) boson field in terms of the original
fields exactly. One simply writes down the most general
three-dimensional
Lagrangian for the effective fields that respects
the symmetries of the original theory. In general,
this effective Lagrangian
contains infinitely many terms -- operator expressions
of arbitrary high dimensions -- and is thus
nonrenormalizable. Therefore, an ultraviolet
cutoff $\Lambda$ is needed. The corresponding
effective coupling parameters $g_{E,i}$ 
($i\!=\!1,2,3$; $E$ stands for effective) are then determined 
by a matching procedure: One computes sufficiently
many static correlation functions, both in the
original and in the effective theory, with as yet
unspecified $g_{E,i}$, and demands that
the results agree at larger distances $R > 1/T$.
The resulting matching relations allow one to express
the effective $g_{E,i}$'s in terms of the
original coupling parameter $g$, $T$, and the
cutoff $\Lambda$:
\begin{equation}
g_{E,i} = g_{E,i}(g,T,\Lambda) \ .
\label{gEi}
\end{equation}
The cutoff $\Lambda$ ($\sim T\!\sim\!2 \pi T$),
also called the factorization scale, is
roughly the momentum below which the effective
theory should take over. Relations (\ref{gEi})
can be understood as perturbation series in the
coupling parameters $g$. Fortunately, the
$g_{E,i}$'s corresponding to interaction operators
with higher dimensions are of higher order in $g$.
Therefore, for calculation at a given order
only a very restricted number of $g_{E,i}$'s
has to be taken into account.

The effective theory approach has been applied by
Braaten and Nieto to 
$\phi^4$ theory \cite{Braaten:1995cm} 
and to QCD \cite{Braaten:1996ju}.

In the case of $g^2 \phi^4/4!$ there is only
one cutoff $\Lambda$ separating the long- and short-distance
regimes \cite{Braaten:1995cm}. Consequently, the free energy 
density $F$ consists of two parts
\begin{equation}
F = F_{\rm S} + F_{\rm L} \ ,
\label{Fphi4}
\end{equation}
where the long-range part $F_{\rm L}$ 
is determined by the effective theory
and contains the effective coupling parameters $g_{E,i}$.
If these $g_{E,i}$'s are expanded in powers of $g$, then
Eq.~(\ref{Fphi4}) yields the formula for $F$ that
was originally obtained perturbatively by
including the ring-diagram resummation
\cite{FrenkelParwani,ArnoldZhai}. It is a
check of consistency of the effective field
approach to show that the total expression for 
$F$ is independent of the factorization scale $\Lambda$.

In QCD one can separate from the short-distance regime
two long-distance regimes (chromoelectric and chromomagnetic)
\cite{Braaten:1996ju} corresponding to two types of screening. 
Therefore one can apply two effective theories.
One is called the electrostatic QCD (EQCD), and contains
electrostatic and magnetostatic effective gauge fields.
It describes the physics of the quark-gluon system
at distances $r\stackrel{>}{\sim}1/m_{\rm E}$, where
$m_{\rm E}\!\sim\!g T$ is the mass scale of
the chromomagnetic screening.
The other is called magnetostatic QCD (MQCD), and contains
only magnetostatic fields. It acts at distances
$r\stackrel{>}{\sim}1/m_{\rm M}$, where
$m_{\rm M} \sim g^2 T$ 
is the magnetic screening mass, which, however, cannot be
calculated perturbatively because of severe infrared
divergences at the order $g^6$ \cite{Linde:px,Gross:1980br}. 
As a consequence, the
free energy density $F$ of hot QCD consists of three
contributions: 
\begin{enumerate}
\item 
the short-distance [$r \stackrel{<}{\sim} 1/(\pi T)$]
contribution $F_{\rm E}$ of nonzero modes, i.e., the modes with
frequencies equal to and higher than the first Matsubara
frequency $\omega_1\!=\!2 \pi T$;
\item
the long-distance contribution $F_{\rm M}$, with
$r \sim 1/m_{\rm E} \sim 1/(g T)$, i.e.,
the collective modes described by EQCD effective theory; and
\item
the "rest" contribution $F_R$ from even larger scales
$r \sim 1/m_{\rm M} \sim 1/(g^2 T)$ of MQCD effective theory.
\end{enumerate}
Each of these three contributions is expected to
be physical, i.e., $\mu$-independent.

\section{The case of QCD}
\label{sec:QCD}

\subsection{Formulas for the approach}
\label{subsec:QCD1}

We will first apply our approach to QCD since this
is physically and experimentally the most interesting
case. The free energy density $F$, for an arbitrary number
$n_f$ of quark generations, has been calculated in
thermal perturbation theory up to terms $\sim\!a^{5/2}$
\cite{ArnoldZhai,Braaten:1996ju}, within
the ${\overline {\rm MS}}$ renormalization scheme.
As mentioned before, the free energy density can be
decomposed into three physically distinct parts
\cite{Braaten:1996ju}
\begin{equation}
F = F_{\rm E} + F_{\rm M} + F_{\rm R} \ .
\label{F3parts}
\end{equation}
Up to order $a^{5/2}$, 
only $F_{\rm E}$ and $F_{\rm M}$ contribute, 
and they are, in principle, perturbatively computable. 
These are the contributions from
the energy intervals $(\Lambda_{\rm E}, \infty)$ and
$(\Lambda_{\rm M}, \Lambda_{\rm E})$, respectively, where the
factorization scales $\Lambda_{\rm E}$ and $\Lambda_{\rm M}$
delimit the ultraviolet (UV) and infrared (IR)
bounds of EQCD and satisfy the inequalities
\begin{equation}
m_{\rm M} (\sim\!g_s^2 T) \ 
\stackrel{<}{\sim} \ \Lambda_{\rm M} \
\stackrel{<}{\sim} \ m_{\rm E} (\sim\!g_s T) \
\stackrel{<}{\sim} \ \Lambda_{\rm E} \
\stackrel{<}{\sim} \ \omega_1 (=\!2 \pi T) .
\label{hierar}
\end{equation}
Apart from fitting into this hierarchy, the factorization scales
$\Lambda_{\rm E}$ and $\Lambda_{\rm M}$ are in principle arbitrary.
The sum of the three contributions is independent of
these scales, and $F_{\rm E}\!+\!F_{\rm M}$ 
is independent of $\Lambda_{\rm E}$.
Once $\Lambda_{\rm E}$ and $\Lambda_{\rm M}$ are fixed, each of the
three contributions is a (quasi)observable, i.e.,
a quantity that is independent of the renormalization
scale and scheme. 
We will show that the
specific RScl-dependence of the available TPS's
for $F_{\rm E}$, $m_{\rm E}$, and $F_{\rm M}$
is in fact consistent with RScl-independence of the full
quantities $F_{\rm E}$, $m_{\rm E}$, and $F_{\rm M}$. 

Specifically, we have for $F_{\rm E}$ the
following TPS \cite{Braaten:1996ju}:
\begin{equation}
F_{\rm E} 
=  - \frac{8 \pi^2}{45} \; T^4 \left[ 
\left( 1\!+\!\frac{21}{32} \; n_f \right) 
- \frac{15}{4} \left(1\!+\!\frac{5}{12} \; n_f \right) 
{\tilde F}_{\rm E}(\Lambda_{\rm E}) \right] \ ,
\label{FE1}
\end{equation}
where the first term represents the contribution of the free
(ideal) quark-gluon gas
\begin{equation}
F_{\rm ideal} =  - \frac{8 \pi^2}{45} \; T^4 
\left( 1\!+\!\frac{21}{32} \; n_f \right) \ ,
\label{Fideal}
\end{equation}
and the ``canonical'' QCD-part ${\tilde F}_{\rm E}(\Lambda_{\rm E})$ is
\begin{equation}
{\tilde F}_{\rm E}(\Lambda_{\rm E}) =
a(\mu^2) \left\{ 1 + \left[
Q(n_f; \Lambda_{\rm E}) + 
L(n_f) \ln \left( \frac{\mu^2}{4 \pi^2 T^2} \right)
\right] a(\mu^2) \right\} + {\cal {O}}(a^3) \ ,
\label{tFE1}
\end{equation}
with
\begin{eqnarray}
Q(n_f; \Lambda_{\rm E}) &=&  
- 18 \frac{(1\!+\!n_f/6)}{(1\!+\!5 \; n_f/12)} \ln \left(
\frac{ {\Lambda}^2_{\rm E} }{ 4 \pi^2 T^2 } \right)
+ \frac{4}{15} 
\frac{(-244.898\!-\!17.2419 \; n_f\!+\!0.415029 \; n_f^2)}
{(1\!+\!5 \; n_f/12)} 
\ ,
\label{Qnf}
\\
L(n_f) & = & \frac{11}{4} \left( 1 - \frac{2 \; n_f}{33} \right) \ .
\label{Lnf}
\end{eqnarray}
Here,\footnote{
We note that there are two misprints in 
a formula for $F_{\rm E}$ ($f_{\rm E} \equiv F_{\rm E}/T$)
in Refs.~\cite{Braaten:1996ju} -- in Eq.~(7) (PRL)
and Eq.~(54) (PRD) -- in the sign of the 
coefficient at $\ln [\Lambda_{\rm E}/(2 \pi T)]$
and in the sign of the term $17.24 \; n_f$.}
$n_f$ is the number of active quark flavors;
$a(\mu^2) \equiv \alpha_s(\mu^2; {\overline {\rm MS}})/\pi
\equiv g_s^2(\mu^2;{\overline {\rm MS}})/(4 \pi^2)$.
The RScl $\mu$ is usually chosen as $\sim\!\pi T$.
We note that $L(n_f)$, the coefficient of the
($\ln \mu^2$)-dependent part at order $a^2$, is exactly
equal to $\beta_0$, the one-loop coefficient
of the QCD beta function of the RGE (\ref{RGE1}). 
It is then exactly this fact which guarantees that the derivative
$\partial {\tilde F}_{\rm E}/ \partial \ln \mu^2$ has
the terms of $\sim\!a^2$ canceled due to the RGE.
Therefore, this variation is $\sim\!a^3$, which is
of the order of the first unknown term in the TPS of
${\tilde F}_{\rm E}$. This shows that the TPS behavior of
${\tilde F}_{\rm E}$ (and $F_{\rm E}$) is compatible with the
supposition of the $\mu$-independence of $F_{\rm E}$.

Further, the available TPS for the 
electric Debye-screening mass $m_{\rm E}$ is
\cite{Braaten:1996ju}
\begin{eqnarray}
{\tilde m}^2_{\rm E} & \equiv &
\frac{1}{4 \pi^2 T^2} \frac{1}{(1\!+\!n_f/6)} \; m^2_{\rm E} 
= a(\mu^2) \left\{ 1 + \left[ P(n_f) +
\beta_0 \ln \left( \frac{\mu^2}{4 \pi^2 T^2} \right) \right] a(\mu^2)
\right\} + {\cal {O}}(a^3) \ ,
\label{mE}
\end{eqnarray}
where 
\begin{equation}
P(n_f) = \left( 0.612377 - 0.488058 \; n_f - 0.0427979 \; n_f^2 \right)/
(1 + n_f/6) \ .
\label{Pnf}
\end{equation}
Again, it is possible to see that 
$\partial {\tilde m}_{\rm E}/\partial \ln \mu^2 \sim a^3$,
compatible with the $\mu$-independence of this quantity.

Let us now consider the part $F_{\rm M}$, which includes
all long-distance contributions, i.e., all
ring-diagram resummation effects. 
The available TPS for $F_{\rm M}$ was calculated
within EQCD (with effective mass $m_{\rm E}$ and
effective coupling $g_{\rm E}$) in Ref.~\cite{Braaten:1996ju}.
Since $g_{\rm E}^2(\mu) = g_s^2(\mu) T \; [ 1\!+\!{\cal O}(g^2)]$,
one can express the result immediately in powers of $g_s$. This gives
\begin{equation}
F_{\rm M} =  \frac{2}{3 \pi} T m_{\rm E}^3 \; 
{\tilde F}_{\rm M}({\Lambda}_{\rm E}; {\Lambda}_{\rm M}) \ ,
\label{FM1}
\end{equation}
with 
\begin{eqnarray}
{\tilde F}_{\rm M}(\Lambda_{\rm E}; {\Lambda}_{\rm M}) &=&
-1 + \left( 0.255838 + \frac{9}{4} 
\ln \frac{ {\Lambda}^2_{\rm E} }{m^2_{\rm E}} \right)
\frac{1}{2 \pi} (1\!+\!n_f/6)^{-1/2} g_s(\mu^2)
\nonumber\\
&& + 27.5569 \frac{1}{4 \pi^2} (1\!+\!n_f/6)^{-1} g_s^2(\mu^2)
+ {\cal {O}}(g^3) \ .
\label{FMexp}
\end{eqnarray}
The $\Lambda_{\rm M}$-dependence
would first show up at terms of higher order $\sim\!g_s^3$. The 
$\Lambda_{\rm E}$-dependence cancels in the sum
$F_{\rm E}\!+\!F_{\rm M}$ to the orders available in 
Eqs.~(\ref{FE1})-(\ref{Lnf}) and (\ref{FM1})-(\ref{FMexp}). 
Equation (\ref{FMexp}) is a series in 
$g_s(\mu^2) \equiv 2 \pi \sqrt{a(\mu^2)}$.
To the order available, it is automatically compatible
with $\mu$-independence of $F_{\rm M}$ 
since $\partial g_s(\mu^2)/ \partial \ln \mu^2 \sim g_s^2$
by the RGE (\ref{RGE1}).

When the screening mass $m_{\rm E}$ (\ref{mE}) is expanded in powers
of $g_s(\mu)$ and inserted in Eqs.~(\ref{FM1})-(\ref{FMexp}), 
a power expansion of $F_{\rm M}$ in $g_s(\mu)$ can be obtained,
starting with $g_s^3$ ($\sim\!m_{\rm E}^3$)
\begin{eqnarray}
F_{\rm M} &=& \frac{2}{3 \pi} k_f^3 T^4 {\Big\{} - g_s^3(\mu^2) +
\frac{1}{2 \pi k_f} \left[0.255838 - 
\frac{9}{2} \ln \left( g_s(\mu^2) \; k_f \right)
+ \frac{9}{2} \ln \left( \Lambda_{\rm E}/T \right) \right] g_s^4(\mu^2)
\nonumber\\
&& + \frac{1}{4 \pi^2 k_f^2} \left[ - \frac{3}{2} 
\left( P(n_f) + \beta_0 \ln \left( \frac{\mu^2}{4 \pi^2 T^2} \right)
\right) k_f^2 + 27.5569 \right] g_s^5(\mu^2) + {\cal O}(g_s^6 \ln g_s)
{\Big \}} \ ,
\label{FMexp2}
\end{eqnarray}
where we used the notation $k_f = (1\!+\!n_f/6)^{1/2}$ and 
Eq.~(\ref{Pnf}).   
Adding it to the expansion (\ref{FE1})-(\ref{tFE1}) for 
$F_{\rm E}$, a TPS for $F_{\rm E+M}$ in powers of
$g_s(\mu)$ up to $\sim\!g_s^5$
is obtained as given in Ref.~\cite{ArnoldZhai} (third entry)
and in Ref.~\cite{Braaten:1996ju} (second entry), i.e.,
a TPS of the form (\ref{genericF}). Interestingly, due to the use
of the expansion of $m_{\rm E}$, the coefficient $C_4(\mu)$
becomes $g_s$-dependent [dependent on $\ln g_s^2(\mu)$],
which represents an additional, although possibly only formal, 
obstacle for the direct application of PA's to such a TPS.\footnote{
Formally, PA's are constructed for TPS's where the coefficients
are independent of the expansion parameter.}
We avoid this problem by using in the free energy
the Pad\'e-resummed squared screening mass $m^2_{\rm E}$.

It is well known that the TPS (\ref{FMexp}) for 
${\tilde F}_{\rm M}$ in QCD 
has very bad divergent behavior, and this is the case to a somewhat 
lesser extent for the TPS of $m_{\rm E}$.
Thus, their direct evaluations do not yield useful predictions.
Of course, this bad divergent behavior is transported
to $F_{\rm M}$ and to $F = F_{\rm E}\!+\!F_{\rm M}$.
To remedy this, the authors of 
Refs.~\cite{Hatsuda:1997wf,Kastening:1997rg}
presented evaluations of $F$ via various
Pad\'e approximants (PA's), which were based on
the TPS of the expansion of $F$ in powers of $g_s$, 
i.e., the sum of Eqs.~(\ref{FE1})-(\ref{tFE1}) and (\ref{FMexp2}),
and no separation was performed. 
Although their results showed significantly
reduced $\mu$-dependence of the Pad\'e resummed values of
$F_{E+M}$ in comparison to the $\mu$-dependence of the
TPS of $F_{E+M}$, we believe that this approach is not
well motivated. 
This is so because it probably leads to partial
double-counting or other interference-like inconsistencies,
as argued in the previous sections.
We will illustrate this argument with 
the following simple example. Suppose that we want to
resum, by a PA, the sum $S \equiv (S_1\!+\!S_2)$
of two observables $S_1$, $S_2$. Suppose that we have
these two observables available as power series of $a$ up to 
next-to-leading order (NLO):
\begin{eqnarray}
S_j &=& a (1 + r_1^{(j)} a) + {\cal {O}}(a^3) \quad (j\!=\!1,2)
\label{Sjs}
\\
\Rightarrow \ S & = & 
2~a \left[ 1 + \frac{1}{2}(r_1^{(1)}\!+\!r_1^{(2)}) a
\right] + {\cal {O}}(a^3) \ .
\label{Ssum}
\end{eqnarray}
Applying to this TPS of the sum a PA,
say [1/1], and expanding the resummed result back in
powers of the coupling $a$, we obtain
\begin{eqnarray}
\lefteqn{
S^{[1/1]} =  2~a \left[ 1 - \frac{1}{2}(r_1^{(1)}\!+\!r_1^{(2)}) a
\right]^{-1}}
\label{SP11}
\\
& & = 2~a \left[ 1 + \frac{1}{2}(r_1^{(1)}\!+\!r_1^{(2)}) a
+ \frac{1}{4} ( r_1^{(1)2} + r_1^{(2)2} + 2 r_1^{(1)} r_1^{(2)} ) a^2 
\right] + {\cal {O}}(a^4) \ .
\label{SP11exp}
\end{eqnarray}
The coefficient at $a^3$ of the power expansion of the
result has a term $2 r_1^{(1)} r_1^{(2)}$, indicating
that the result contains mixing effects at $\sim\!a^3$.
This indicates some kind of interference effect
between the two amplitudes for the observables (processes)
$S_1$ and $S_2$. This is not acceptable because $S$ is
the sum of these two independent observables, i.e.,
the two contributions should be summed up incoherently.
This argument remains basically unchanged when
different PA's are applied, and/or when the TPS's are 
of higher order, either in $a$ or in $g_s$.

It is true that expansion of $F_{E+M}$ in powers of
$g_s$ (using also expansion of $m_{\rm E}$ in $g_s$)
gives us a TPS of relatively high order ($\sim g_s^5 \sim a^{5/2}$),
and that we can thus apply PA's of relatively high order.
Further, the higher order PA's are known to possess in
general a weaker $\mu$-dependence than the corresponding
TPS \cite{Gardi:1996iq}. However, such an
approach would predict nonphysical higher order
effects and thus lead to unreliable predictions
for the sum ($F_{E+M}$).

As argued in the previous sections, we follow a more
conservative approach, by summing up by PA's
each TPS (for $F_{\rm E}$, $m_{\rm E}$, $F_{\rm M}$) separately.
This approach gives us an additional freedom --
to choose the RScl in each of these TPS's separately,
in the natural range $\mu\!\sim\!1/r$ where $r$ is a typical
distance associated with each observable. For TPS's
of $F_{\rm E}$, $m_{\rm E}$, $F_{\rm M}$ this implies the RScl choices:
$\mu_{\rm E}\!\sim\!\omega_1$ ($=\!2 \pi T$);
$\mu_m\!\sim\!m_{\rm E}$; and $\mu_{\rm M}\!\sim\!m_{\rm E}$.
Although, in this approach, we have to take as bases
the TPS's (\ref{tFE1}), (\ref{mE}), (\ref{FMexp})
of very low order (NLO), and some of these TPS's show
very divergent behavior, we will now show 
that the results nonetheless show remarkably weak dependence
on $\mu$ and $\Lambda_{\rm E}$.

\subsection{Numerical results}
\label{subsec:QCD2}

In our numerical analysis we used, unless otherwise stated,
for the number of active (massless) quark flavors $n_f\!=\!3$, 
and for the QCD coupling constant the reference value
$\alpha_s(M_{\tau}^2, {\overline {\rm MS}}) = 0.334$.
We used the ${\overline {\rm MS}}$ scheme, and 
for the $\beta$-function we used, for definiteness,
the $[2/3]$ PA.\footnote{
As argued in Refs.~\cite{CK1},
this PA gives us a reasonable quasianalytic continuation of
the ${\overline {\rm MS}}$ $\beta$-function for
values of $\alpha_s(\mu)$ up to 
$\alpha_s(\mu) = \pi \times 0.32 \approx 1.0$,
i.e., for values where the TPS $\beta$ gives highly unreliable results.}

\begin{figure}[htb]
\setlength{\unitlength}{1.cm}
\begin{center}
\epsfig{file=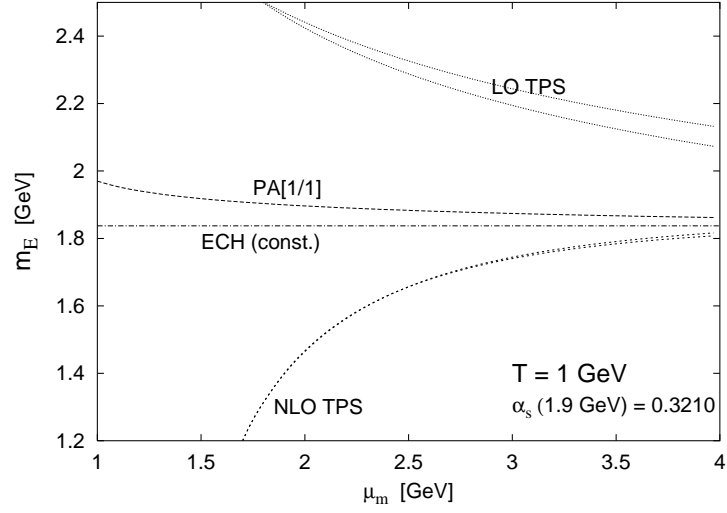, width=10.cm}
\end{center}
\vspace{0.0cm}
\caption{\footnotesize The Debye screening mass $m_{\rm E}$ as function
of the renormalization scale (RScl) $\mu_m$, when $T\!=\!1$ GeV.
The upper of the two curves LO TPS and NLO TPS, respectively,
has $a(\mu_m^2)$ evolved by the one-loop and two-loop
RGE from $a(m_{\tau}^2)$. All the other curves have
$a(\mu_m^2)$ evolved by the four-loop PA $[2/3]$ beta function.} 
\label{mEvsmuT=1}
\end{figure}

In Fig.~\ref{mEvsmuT=1} we present the results for the
screening mass $m_{\rm E}$ as 
a function of the corresponding RScl $\mu_m$,
for the temperature choice $T\!=\!1$ GeV.
All approximations are based on the (NLO) TPS (\ref{mE}).
We see that the diagonal PA $[1/1](a)$ significantly 
reduces the RScl-dependence, in comparison to the
LO and NLO TPS's, in accordance with the arguments of
Ref.~\cite{Gardi:1996iq}. Furthermore, the effective charge
(ECH) method \cite{ECH,KKP,Gupta} of fixing the
RScl in the NLO TPS gives us a value (fixed by definition)
not far from the PA [1/1] values.\footnote{
A generalization of the method of diagonal PA's has been
developed \cite{CK2,CK1}, which gives complete RScl-independence
\cite{CK2}, or RScl- and RSch-independence \cite{CK1}.
However, we have here TPS's available only at the
low NLO order, at which the aforementioned
approximants basically reduce to those of the ECH method.}
We will choose the ``physical'' screening mass to be determined
by the condition $m_{E}^{[1/1]}(\mu_m)\!=\!\mu_m$ 
($\equiv\!m_{\rm E}^{(0)}$), which
gives us the value $m_{\rm E}^{(0)}\!=\!1.9$ GeV for $T\!=\!1$ GeV.

\begin{figure}[htb]
\begin{minipage}[b]{.49\linewidth}
 \centering\epsfig{file=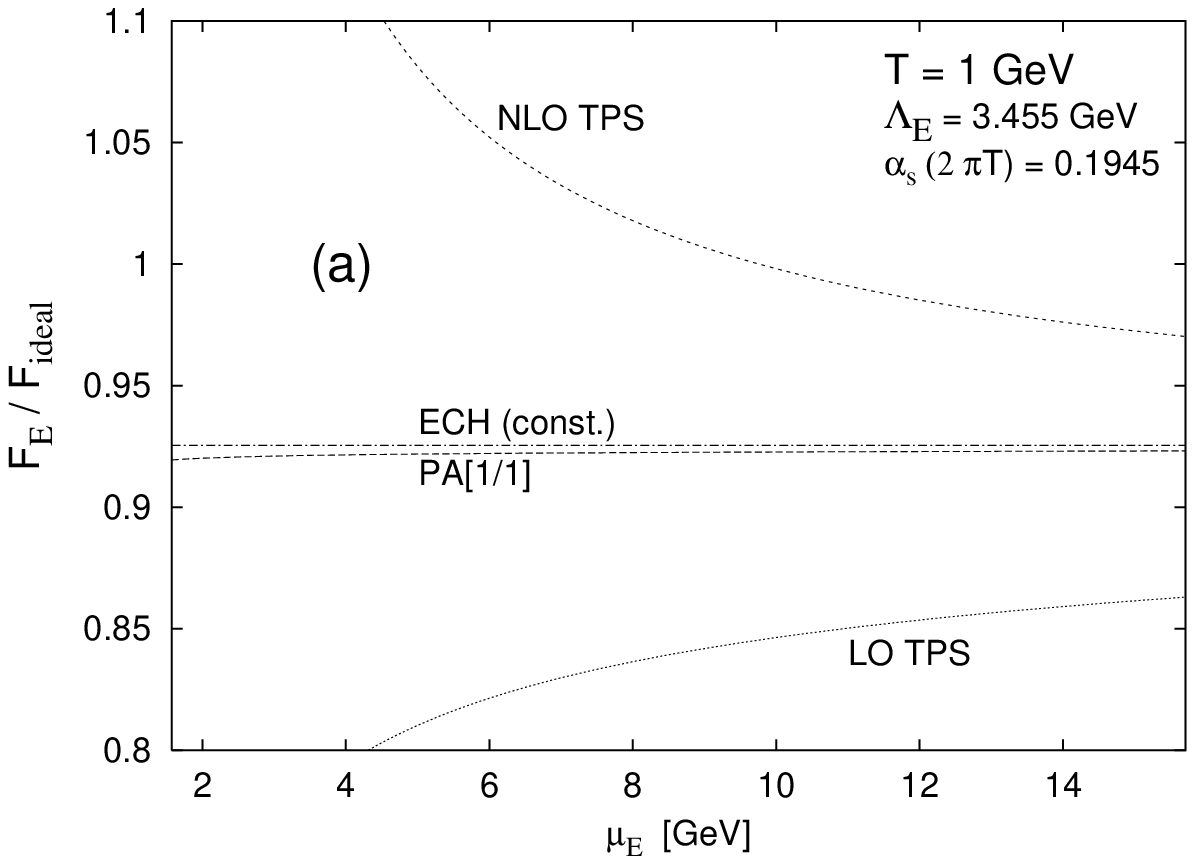,width=\linewidth}
\end{minipage}
\begin{minipage}[b]{.49\linewidth}
 \centering\epsfig{file=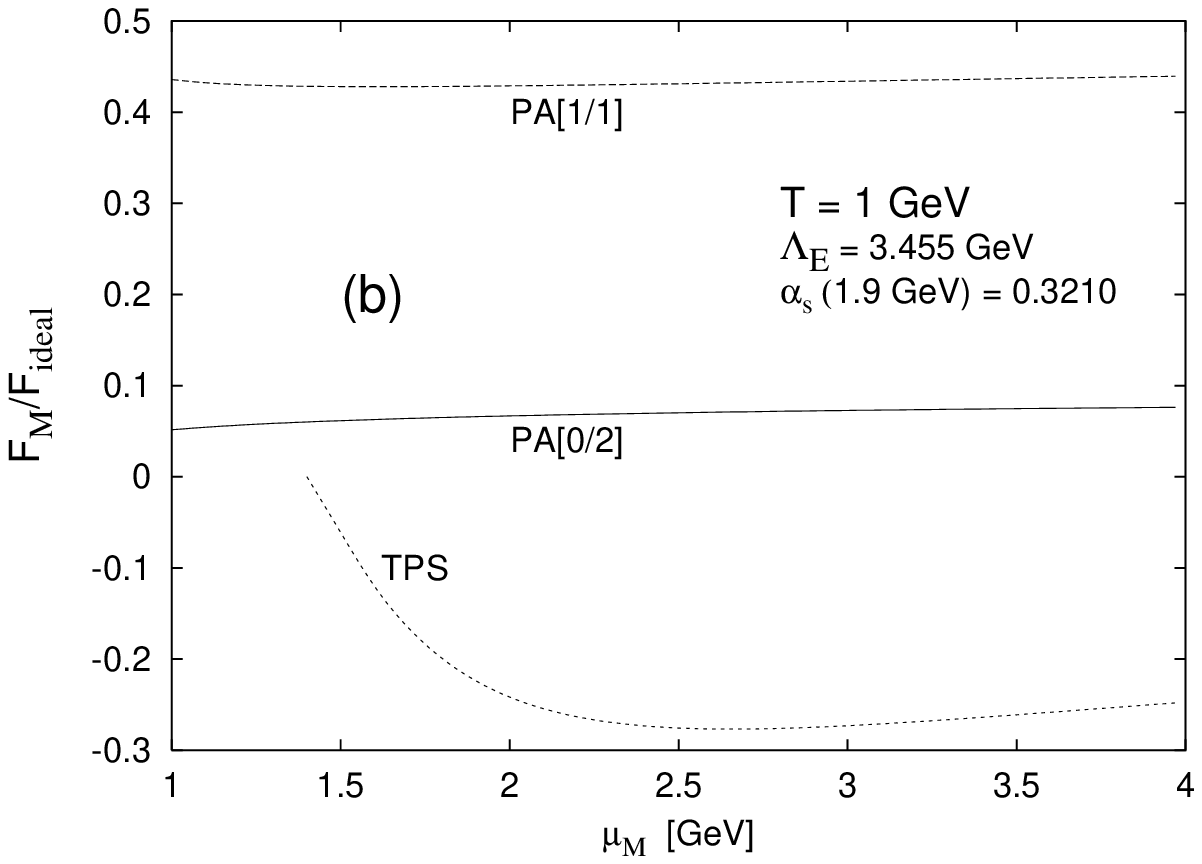,width=\linewidth}
\end{minipage}
\vspace{0.2cm}
\caption{\footnotesize The $F_{\rm E}$ (a) and $F_{\rm M}$ (b) contributions
to the free energy $F$, as functions of the corresponding
RScl $\mu_{\rm E}$, $\mu_{\rm M}$.}
\label{FEMvsmuT=1}
\end{figure}

In Figs.~\ref{FEMvsmuT=1}(a) and \ref{FEMvsmuT=1}(b), 
we present the results
for $F_{\rm E}$ and $F_{\rm M}$ as functions
of the respective RScl's $\mu_{\rm E}$ and $\mu_{\rm M}$.
The factorization scale $\Lambda_{\rm E}$ was chosen,
in the spirit of the hierarchy relations (\ref{hierar}),
to be the logarithmic mean of the typical
scales $\mu_{\rm E} \sim 2 \pi T$ and $\mu_{\rm M} \sim m_{\rm E}$:
$\Lambda_{\rm E} = \sqrt{ 2 \pi T m_{\rm E}^{(0)}}$, 
where $m_{\rm E}^{(0)}\!=\!1.9$ GeV
is the aforementioned ``physical'' $[1/1](a)$ screening mass.
The PA $[1/1](a)$ for $F_{\rm E}$ is
based on the NLO TPS (\ref{FE1})-(\ref{tFE1}).
We see that the PA [1/1] for $F_{\rm E}$ has drastically reduced
the RScl-dependence, and its values are very close to those of the 
ECH prediction.
 
On the other hand, the situation with $F_{\rm M}$ is less
favorable, as seen from Fig.~\ref{FEMvsmuT=1} (b).
The two PA's $[1/1](g_s)$ and $[0/2](g_s)$ were constructed from the
TPS in powers of $g_s$ (not $a$) of Eq.~(\ref{FMexp}), 
taking for $m_{\rm E}$, appearing in Eqs.~(\ref{FM1}) and
(\ref{FMexp}), the PA $[1/1](a)$ with the RScl's for $m_{\rm E}$
and $F_{\rm M}$ taken equal: $\mu_m\!=\!\mu_{\rm M}$. In the case of
the (NLO) TPS result for $F_{\rm M}$, we used for $m_{\rm E}$ the NLO TPS
result as well. While the two PA's give reduced
RScl-dependence of $F_{\rm M}$, their values differ
drastically. In the case of the PA summation of
$F_{\rm E}$ and $m_{\rm E}$ the diagonal PA's $[1/1](a)$ are more
physically motivated than the off-diagonal ones
$[0/2](a)$, due to the RScl-independence of $[1/1](a(\mu))$
in the large-$\beta_0$ approximation \cite{Gardi:1996iq}.
That is why we did not employ the $[0/2](a)$ PA's for
$F_{\rm E}$ and $m_{\rm E}$. On the other hand, for $F_{\rm M}$ we
have a TPS in $g_s(\mu)=2 \pi \sqrt{a(\mu)}$ of
Eq.~(\ref{FMexp}), and not in $a(\mu)$, and
therefore the choice $[1/1](g_s)$ for the right-hand side (RHS) 
of Eq.~(\ref{FMexp}) is not physically better motivated
than the choice $[0/2](g_s)$. This can be seen also from
the comparably weak RScl-dependence of both
PA approximants in Fig.~\ref{FEMvsmuT=1}(b). In order
to choose between the two, we have to study, in addition,
their variation under the variation of the
factorization scale $\Lambda_{\rm E}$. This will be discussed
just below.

\begin{figure}[htb]
\begin{minipage}[b]{.49\linewidth}
 \centering\epsfig{file=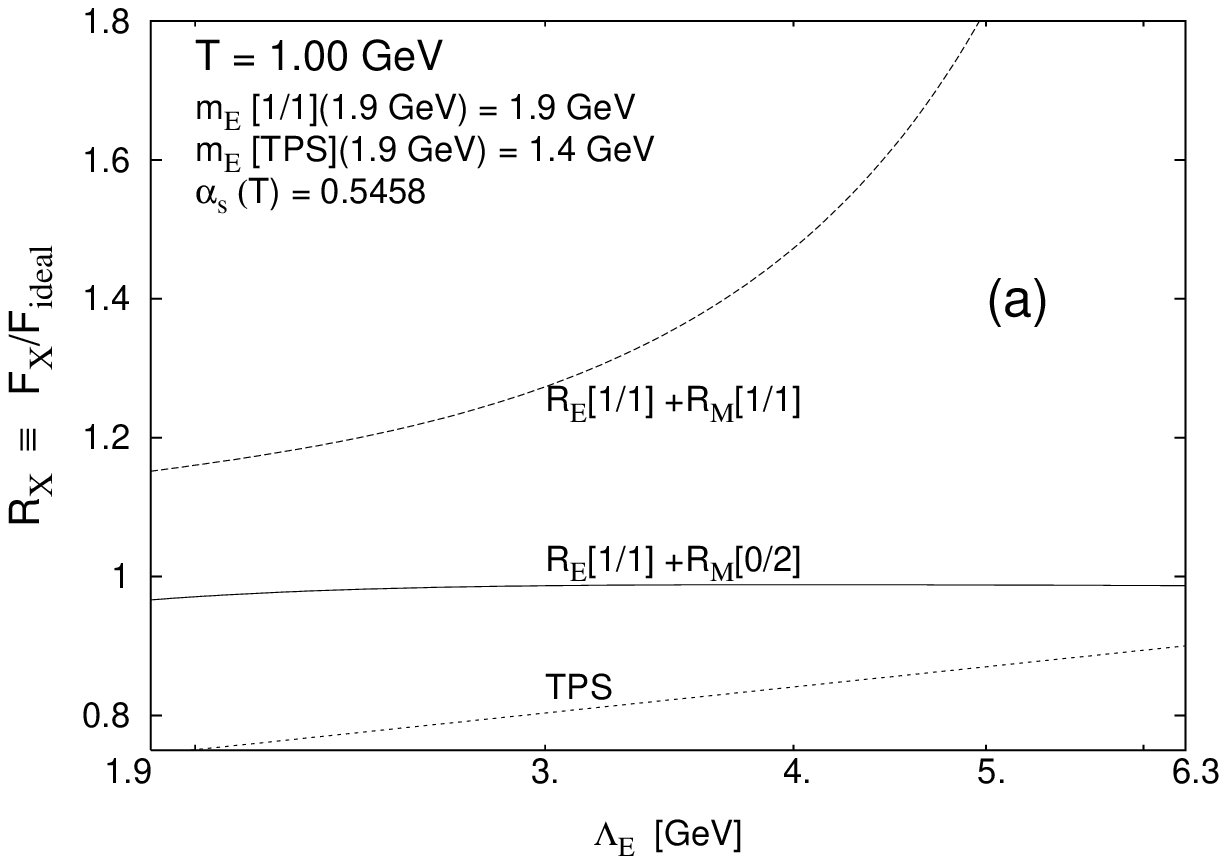,width=\linewidth}
\end{minipage}
\begin{minipage}[b]{.49\linewidth}
 \centering\epsfig{file=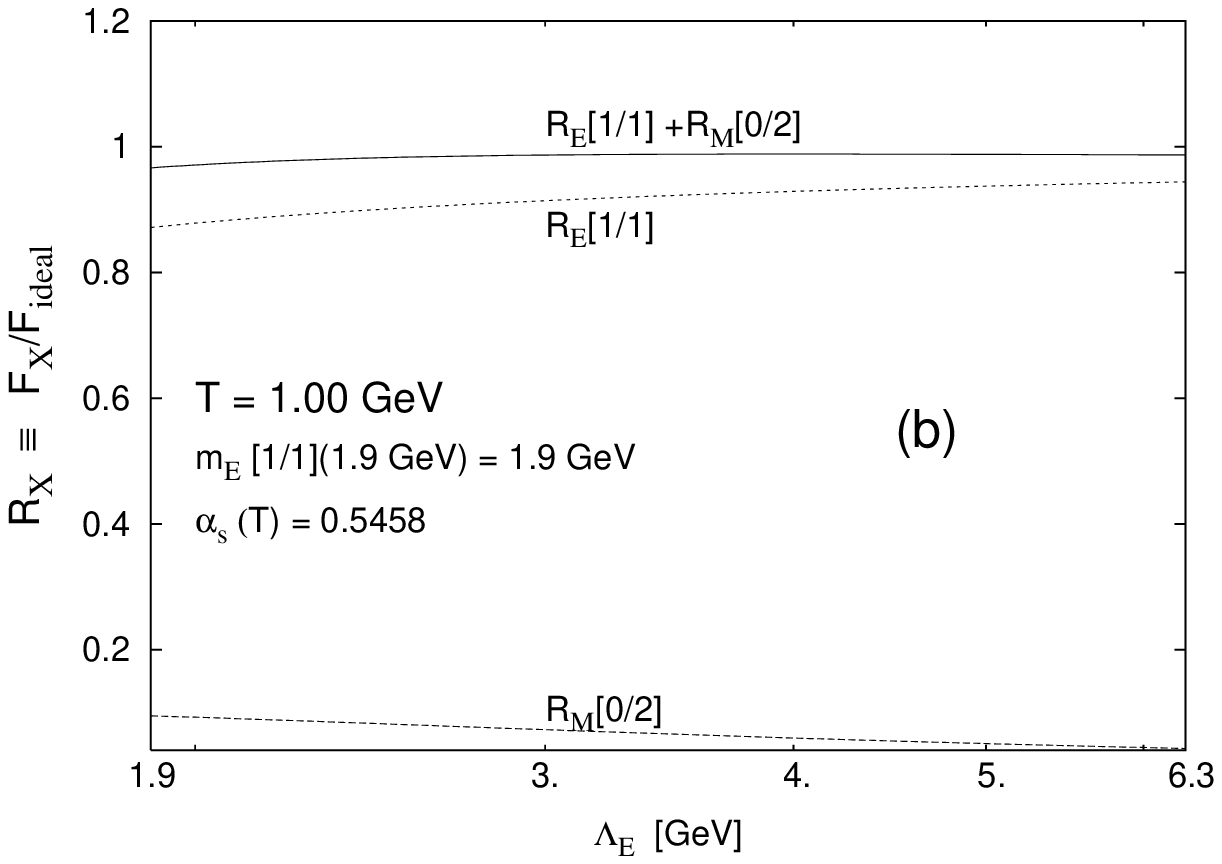,width=\linewidth}
\end{minipage}
\vspace{0.2cm}
\caption{\footnotesize 
(a) Various approximants for the normalized
sum $R_{E+M}\!\equiv\!F_{E+M}/F_{\rm ideal}$ (a) as functions of the
factorization scale $\Lambda_{\rm E}$. (b) shows the
separate ($E$ and $M$) contributions for the stable
approximation $R_{E}[1/1]\!+\!R_{M}[0/2]$. Other details
are explained in the text.}
\label{FEMvsLET=1}
\end{figure}

In Figs.~\ref{FEMvsLET=1}(a) and \ref{FEMvsLET=1}(b), 
we present the $\Lambda_{\rm E}$-dependence
of the approximants.
The RScl's were fixed, 
in accordance with the hierarchies (\ref{hierar}),
in the following way: $\mu_{\rm E}\!=\!2 \pi T$, 
$\mu_{\rm M}\!=\!\mu_m\!=\!m_{\rm E}$, where $m_{\rm E}$ has the
aforementioned ``physical'' value: 
$m_{\rm E}\!=\!m_{\rm E}^{(0)}\!=\!1.9$ GeV, for the case of the PA
approximants and $1.4$ GeV for the case of the TPS's.
Figure \ref{FEMvsLET=1}(a) shows that the choice $[1/1](g_s)$ for the
TPS of Eq.~(\ref{FMexp}) in $F_{\rm M}$
leads to strong, and thus unphysical, $\Lambda_{\rm E}$-dependence
of $F_{E+M}$. Thus we have to discard the $[1/1](g_s)$
result of the TPS (\ref{FMexp}). The choice
$[0/2](g_s)$ for this part leads, on the other hand,
to a result for $F_{E+M}$ which is remarkably
stable under variation of $\Lambda_{\rm E}$ in the
entire interval $m_{\rm E}^{(0)} \leq \Lambda_{\rm E} \leq 2 \pi T$.

Therefore, we conclude that the choice
$[1/1](a)$ for $F_{\rm E}$ and for $m_{\rm E}$,
and the choice $[0/2](g_s)$ for the TPS of Eq.~(\ref{FMexp}),
i.e., the curve $F_{\rm E}[1/1]\!+\!F_{\rm M}[0/2]$ in
Fig.~\ref{FEMvsLET=1}(a), is the least unreliable among all the curves.

The aforementioned properties
of various approximants under changes of the
RScl's and of $\Lambda_{\rm E}$ remain qualitatively the same
when we change the value of the temperature.
In Figs.~\ref{FEMvsTg}(a) and \ref{FEMvsTg}(b), we present the results
for the approximant $(F_{E}[1/1]\!+\!F_{M}[0/2])/F_{\rm ideal}$
as a function of the temperature $T$ and of
the coupling $g_s(2 \pi T)$ [$=\!2 \pi \sqrt{a(2 \pi T)}$].
The middle curve (mid $\Lambda$)
is for the canonical choice of the scales,
$\mu_{\rm E}\!=\!2 \pi T$ and $\mu_{\rm M}\!=\!\mu_m\!=\!m_{\rm E}$
where $m_{\rm E}$ is, for each $T$, determined in the aforementioned
way: $m_{\rm E}^{[1/1]}(\mu_m)=\mu_m$. The factorization scale
is taken to be $\Lambda_{\rm E}\!=\!\sqrt{\mu_{\rm E} \mu_{\rm M}}$.
The lower curve (low $\Lambda$) is for the
choice of the scales being all at the lower
extreme: $\mu_{\rm E}\!=\!\mu_{\rm M}\!=\!\mu_m\!=\!m_{\rm E}\!=\!\Lambda_{\rm E}$,
where $m_{\rm E}$ is determined in the aforementioned way.
The upper curve (high $\Lambda$) is for the
choice of the scales at the upper extreme:
$\mu_{\rm E}\!=\!\mu_{\rm M}\!=\!\mu_m\!=\!\Lambda_{\rm E}\!=\!2 \pi T$.
We notice that the curve of the low $\Lambda$ choice follows
well the curve for the canonical (mid $\Lambda$) choice
in the entire depicted region of $T$ (of $g_s$).
The other choice of PA's ($[1/1](a)$ for $F_{\rm E}$ and $m_{\rm E}$;
$[1/1](g_s)$ for ${\tilde F}_{\rm M}$ 
of Eq.~(\ref{FMexp})) gives values well
outside this range -- see Fig.~\ref{FEMvsLET=1}(a). 
The TPS's [for $F_{\rm E}$, $m_{\rm E}$, and for 
${\tilde F}_{\rm M}$ of Eq.~(\ref{FMexp})]
do give us values in qualitative agreement with the lattice results
for a specific choice of scales (mid $\Lambda$)
[see Fig.~\ref{FEMvsLET=1}(a)], but the TPS values
change drastically when some of these scales,
in particular $\mu_{\rm M}$, change [see Fig.~\ref{FEMvsmuT=1}(b)].
In Figs.~\ref{FEMvsTg}(a) and \ref{FEMvsTg}(b), we included,
for comparison, the values of the TPS up to ${\cal O}(g_s^5)$ in
powers of $g_s(2 \pi T)$, i.e., the sum of the TPS's
(\ref{FE1})-(\ref{tFE1}) and (\ref{FMexp2}).
We note that the values for the
TPS up to ${\cal O}(g_s^4)$ would be above $1$.

As all the curves up to this point have been given for
the choice of three active massless quark flavors ($n_f\!=\!3$),
we present in Figs.~\ref{FEMvsTgnf}(a) and \ref{FEMvsTgnf}(b), the
results for $n_f\!=\!0,2,3,4,6$ (and for the mid $\Lambda$
choice of the scales). We see that the curve for $n_f\!=\!4$
differs little from that for $n_f\!=\!3$ in most of the
covered parameter space, while the curve for $n_f\!=\!0$
differs significantly. However, in the temperature region
of particular interest ($T \sim 0.1$-10 GeV), the
choices $n_f\!=\!3$ or 4 are expected to be more
realistic.

\subsection{Comparison with other approaches}
\label{subsec:QCD3}

We will now compare the results presented in 
Figs.~\ref{FEMvsTg} and \ref{FEMvsTgnf} with 
those of some other approaches.

As mentioned in Sec.~\ref{subsec:QCD1},
Kastening \cite{Kastening:1997rg} and
Hatsuda \cite{Hatsuda:1997wf} applied
high order PA's to the full TPS for
the sum $R \equiv (F_{\rm E} + F_{\rm M})/F_{\rm ideal}$,
i.e.,\footnote{
We note that in the thermodynamic limit
$F\!=\!- p$, where $p$ is pressure. Thus,
$R\!=\!p/p_{\rm SB}$ where $p_{\rm SB}\!\equiv\!p_{\rm ideal}$
is the ideal gas pressure.}
TPS in powers of $g_s(\mu)$ including $\sim\!g_s^5$.
Kastening \cite{Kastening:1997rg} showed that, in the case of value
$g_s(T) \approx 1.1$ [$\alpha_s(T)\!=\!0.1$, $n_f\!=\!6$], 
the application of PA's
$[2/2](g_s)$ and $[2/3](g_s)$ reduced the RScl dependence 
in comparison with the TPS results. At larger values 
$g_s(2 \pi T) \approx 2$ [$\alpha_s(2 \pi T)\!=\!1/3$,
$n_f\!=\!3$] there was no significant reduction.
On the other hand, Hatsuda \cite{Hatsuda:1997wf}
applied a kind of modified PA's, by postulating 
that the expressions in the numerator and the
denominator of the PA's have zero coefficients
for the term $\sim\!g_s$. The RScl dependence
was then shown to be significantly
reduced for many such modified PA's
($[2/3], [3/2], [2/4], [4/2]$; with $n_f\!=\!4$).
Further, his curves for the $R$ obtained
as a function of $\alpha_s(2 \pi T)$
were qualitatively similar to our
curves in Figs.~\ref{FEMvsTg}(b) and \ref{FEMvsTgnf}(b),
with $R_{\rm min} = 0.97$-$0.98$. However, the differences
between his curves were significant,
and there was no clear principle to choose
any specific one of them. Furthermore, in
our Sec.~\ref{subsec:QCD1} we stressed
a more physical point of criticism of this approach.

A reliable comparison with lattice results is
hampered by the fact that (QCD-)lattice calculations
have reproduced thermodynamic quantities
(free energy or pressure, entropy, etc.) only for
temperatures between the critical temperature
$T_c$ and  $4.5 T_c$ (i.e., for $T \alt 1$ GeV).
In this temperature region the resulting values
for $R$ are less than $0.87$ when $n_f\!=\!0$ \cite{Boyd:1995zg}, 
significantly lower than $1$.
If the number of (massless or light) quark
flavors $n_f$ is larger than zero, the finite cutoff effects are 
not quite under control and are estimated \cite{Karsch:2000ps}
to increase the calculated $R(n_f)$ by about $15 \%$,
giving at $T = 3.5 T_c$ ($\approx 0.7$ GeV) the values
$R \approx 0.90 \pm 0.04$ for $n_f\!=\!2$
(cf.~Fig.~3 of Ref.~\cite{Karsch:2000ps}).
On the other hand, our approach, which is based
on improved perturbative considerations, is
expected to be reliable only for much higher
temperatures $T \agt 10$ GeV where the corresponding
values of the effective coupling
parameters $g_s(2 \pi T)$ are not much higher than $1$.
At such high temperatures we predict $R \approx 0.975$
or higher, for $n_f \leq 4$ [cf.~Fig.~\ref{FEMvsTgnf}(a)],
i.e., very near to $1$. These predictions are
not incompatible with the lattice results, however. 
For example, if we take 
from Fig.~3 of Ref.~\cite{Karsch:2000ps}
the continuum estimate for $p/T^4$ (with $n_f\!=\!2$), 
which is given as a band-curve for $T \leq 3.5 T_c$
($\approx 0.7$ GeV), 
and extrapolate it smoothly to
$T \agt 10$ GeV, we can obtain a fair agreement
with our results.
Further, we note that for $T \agt 10$ GeV,
$g_s(2 \pi T)$ can still be rather high 
[$g_s(2 \pi T)\!=\!1.1$-$1.2$]
and the numerical evaluations of perturbative TPS
for such $g_s(2 \pi T)$ differ from our results significantly.
For example, the TPS up to ${\cal O}(g^5)$, partly visible in
Fig.~\ref{FEMvsTg}, gives us for $g_s(2 \pi T) \approx 1.2$
($1.0$) [$ \Leftrightarrow \ T \approx 10$ ($100$) GeV] 
the value $0.925$ ($0.952$).

One interesting point in connection with
lattice data is the dependence of $R$  on the number
$n_f$ of light or massless quark flavors.
It is well known that the ideal gas expression for $F$
shows an increase for $|F|$ with increasing $n_f$,
Eq.~(\ref{Fideal}). Further, while lattice calculations
show that the ratio
$R(n_f) \equiv F(n_f)/F(n_f)_{\rm ideal}$
also increases with increasing $n_f$
at $T\!\sim\!T_c$, they indicate an inversion
of the $n_f$-dependence of $R$ at the highest 
available $T$ values ($T \approx 4 T_c$) \cite{Karsch:2000ps}.
This is in accordance with our finding 
[see Fig.~\ref{FEMvsTgnf}(a)] that
at $T \gg T_c$ the ratio $R$ decreases with
increasing $n_f$.

A calculation in the $\Phi$-derivable
approximation scheme using hard thermal
loop (HTL) propagators \cite{Peshier:2000hx}
gives lower minimal values of $R$ ($R_{\rm min} \approx 0.82$,
see Fig.~3 there, where $n_f\!=\!0$),
and the minimum is at a much higher value
of $g_s$ ($g_s \approx 2.5$) than
our curves [$g_s(2 \pi T) \approx 1.1$].
However, the $\Phi$-derivable
approximation was performed at the
leading-loop order,
and the expansion of the result $R$ in 
powers of $g_s$ underestimates the
positive $\sim\!g_s^3$ term by a factor
of $1/4$. This indicates that the
correction of this effect would push
the results for $R$ higher.

\section{Resummation results in the case of $\phi^4$ theory} 
\label{sec:resumphi4}

We can apply the same methods of resummation in
the massless scalar $\phi^4$ theory. In this case,
we can use the corresponding TPS results from 
Ref.~\cite{Braaten:1995cm}. The interaction is
\begin{equation}
{\cal L}_{\rm int} = \frac{g^2}{4!} \phi^4 \ ,
\label{intphi4}
\end{equation}
and the $\beta$-function in the ${\overline {\rm MS}}$
scheme is known to five loops \cite{Kleinert:rg}
\begin{equation}
\frac{\partial a}{\partial \ln \mu} =
3 a^2 - \frac{17}{3} a^3 + 32.54 a^4
- 271.6 a^5 + 2848.6 a^6 + {\cal O}(a^7) \ ,
\label{RGEphi4}
\end{equation}
where $a(\mu)\!\equiv\!g^2(\mu)/(16 \pi^2)$.
For definiteness, we choose the PA [3/3] for this $\beta$-function,
in order to simulate better the running in the large-$g(\mu)$
region. The high energy contribution $F_{\rm S}$ 
to the free energy $F$ is 
\begin{eqnarray} 
F_{\rm S}(T) & = & - \frac{\pi^2}{90} \; T^4 
\left( 1 - \frac{5}{4} {\tilde F}_{\rm S}(T)  \right) \ ,
\label{F1}
\\
{\tilde F}_{\rm S}(T) & = & a(\mu^2) \left\{
1 -  \left[ 3 \ln \left( \frac{\mu}{4 \pi T} \right) +
9.29324  \right] a(\mu^2) \right\} + {\cal O}(a^3) \ .
\label{tF1}
\end{eqnarray}
In contrast with the analogous QCD quantity $F_{\rm E}$,
$F_{\rm S}$ does not depend on the factorization scale
$\Lambda_F$ (in QCD: $\Lambda_{\rm E}$) at the available order. 
The TPS for the Debye screening mass $m_L$ is
\begin{eqnarray} 
{\tilde m}_L^2(\Lambda_F) & \equiv &
\frac{3}{2 \pi^2 T^2} m_L^2(\Lambda_F)
\nonumber\\
&=& a(\mu^2) \left\{
1 +  \left[ - 3 \ln \left( \frac{\mu}{4 \pi T} \right) 
+ 4 \ln \left( \frac{\Lambda_F}{4 \pi T} \right)
+ 5.39289 \right]  a(\mu^2) \right\} + {\cal O}(a^3) \ .
\label{mL}
\end{eqnarray}
In contrast with the analogous QCD quantity $m_{\rm E}$,
the mass $m_L$ does depend on the factorization scale.
The low energy contribution $F_{\rm L}$ to $F$ has the following
associated TPS:
\begin{eqnarray}
\frac{12 \pi}{T} \frac{F_{\rm L}}{m_L^3(\Lambda_F)}
&=& -1 + \frac{3}{2} \sqrt{\frac{3}{2}} \frac{1}{4 \pi} g(\mu^2)
 + \left[ 4 \ln \left( \frac{ \Lambda_F }{ 4 m_L(\Lambda_F) } \right)
+ \frac{9}{2} \right] \frac{3}{2} \frac{1}{16 \pi^2} g^2(\mu^2)   
+ {\cal O}(g^3) \ .
\label{F2exp}
\end{eqnarray}
The factorization scale dependence now cancels out
within $F_{\rm L}$, up to the available order
($\sim\!g^5$). We have the hierarchy (\ref{hierar})
as in QCD, with the substitutions $m_{\rm E} \mapsto m_L$ and
$\Lambda_{\rm E} \mapsto \Lambda_F$. 
The entire analysis of the QCD case, as described in
the previous section, can now be repeated following
the same procedures. For the coupling parameter
we chose the reference value $g(2 \pi \ {\rm GeV}) = 4$.
For the temperature choice $T\!=\!1$ GeV,
the results for the screening mass $m_L$ as a function of the 
corresponding RScl $\mu_m$ are presented in Fig.~\ref{mLvsmuT=1},
and for $F_{\rm S}$ and $F_{\rm L}$ as functions of the corresponding
RScl's $\mu_{\rm S}$ and $\mu_{\rm L}$ in Figs.~\ref{F12vsmuT=1}(a)
and \ref{F12vsmuT=1}(b).
The factorization scale $\Lambda_F$ 
was determined similarly as in the
QCD case: $\Lambda_F\!=\!\sqrt{2 \pi T m_L^{(0)}}$,
where in turn $m_L^{(0)}$ is the ``physical''
screening mass determined by the condition
$m_L^{[1/1]}(\mu_m; \Lambda_F) = \mu_m$ ($\equiv m_L^{(0)}$),
which gives us the values $m_L^{(0)}\!=\!0.882$ GeV
and $\Lambda_F\!=\!2.354$ GeV (for $T\!=\!1$ GeV). 
The PA curves for $F_{\rm L}$ in Fig.~\ref{F12vsmuT=1}(b) 
were obtained by applying the corresponding PA's 
$[1/1](g)$ or $[0/2](g)$ to the
RHS of Eq.~(\ref{F2exp}) and employing for 
$m_L$ the PA $[1/1](a)$ at the same RScl ($\mu_m\!=\!\mu_{\rm L}$).
The TPS curve was obtained by using the TPS of the RHS of
Eq.~(\ref{F2exp}), using for $m_L$ the (NLO) TPS
(\ref{mL}). The $[0/2]$ curve in Fig.~\ref{F12vsmuT=1}(b)
is much less RScl-dependent than $[1/1]$, but not
significantly less than the TPS curve.

The dependence of the results on the factorization
scale $\Lambda_F$ is depicted in Fig.~\ref{F12vsLFT=1}.
We see that the PA choice $[1/1](a)$ for $F_{\rm S}$ (and $m_L$)
and $[0/2](g)$ for $F_{\rm L}$ results in weak $\Lambda_F$-dependence
which is comparable with the TPS result,\footnote{
The latter is $\Lambda_F$-dependent because we apply in
Eq.~(\ref{F2exp}) in this case the (NLO) TPS
value for $m_L$.} while the $[1/1](g)$ choice again gives
unacceptably strong $\Lambda_F$-dependence.

When the temperature is changing, so is $g(2 \pi T)$
according to the RGE (\ref{RGEphi4}) (we use [3/3] PA
on the RHS). In Fig.~\ref{F12vsg}(a) we present the
results for the choice $[1/1](a)$ (for $F_{\rm S}$ and $m_L$)
and $[0/2](g)$ [for the RHS of Eq.~(\ref{F2exp})], 
and for the NLO TPS choice
[for $F_{\rm S}$, $m_L$, RHS of Eq.~(\ref{F2exp})]. 
The ``middle'' values (mid $\Lambda$)
are chosen for the relevant scales $\mu_{\rm S}\!=\!2 \pi T$,
$\mu_{\rm L}\!=\!\mu_m\!=\!m_L^{[1/1]}(\mu_m;\Lambda_F)$,
and $\Lambda_F\!=\!\sqrt{\mu_{\rm S} \mu_{\rm L}}$. If these scales
are changed to low $\Lambda$ 
[$\mu_{\rm S}\!=\!\mu_{\rm L}\!=\!\mu_m\!=\!\Lambda_F$ with 
$m_L^{[1/1]}(\mu_m;\Lambda_F)\!=\!\mu_m$],
or to high $\Lambda$ 
($\mu_{\rm S}\!=\!\mu_{\rm L}\!=\!\mu_m\!=\!\Lambda_F\!=\!2 \pi T$),
the PA results vary with $g(2 \pi T)$ 
as presented in Fig.~\ref{F12vsg}(b).

\section{Summary and Conclusions}

We have presented here a method to improve the
predictive power of perturbative thermal field theory.
Predictions obtained by ordinary (truncated)
perturbation theory (including a resummation to get rid
of finite-$T$ infrared divergences) suffer from serious 
divergence and renormalization scale (RScl) ambiguity problems.
Therefore, a careful reorganization of these series
is needed, which converts their physical content into
expressions with better convergence behavior
and greater stability under the variation of the RScl.

Efforts in this direction have been undertaken by
various authors during the last five years
and considerable improvement has been achieved.
However, most of these attempts have
improved either the divergence problem or the
RScl-ambiguity, but not both.

We suggest an approach that addresses both
problems, but concentrating mainly
on the unphysical RScl-dependence. It is based on a
physically motivated separation of the 
free energy density $F$ into
parts which are separately RScl-independent
since each of them has an empirical significance
on its own. This can be achieved because the
separation principle is determined by observable
effects (screening of the different massless
degrees of freedom). In this way one
simultaneously obtains a clear separation
between terms which include
(ring-diagram) resummation effects
on the one hand, and purely perturbative
contributions on the other hand, which are free
of any such resummation. This is gratifying
since it allows us to consistently apply
Pad\'e approximants (PA's) to the available
truncated perturbation series (TPS's) 
of the individual parts,
thereby avoiding the danger of double counting
and spurious interferences of contributions
from different kinematical regimes. We also
Pad\'e resum the (observable) screening mass which appears
in the long-distance part of the free
energy density. The resulting expression
for the free energy density has strongly reduced  
RScl-dependence. In addition, we expect this method
to show a good convergence when applied to
TPS's of higher order, based on the well-known
behavior of PA's \cite{Padebook}.

Due to its close connection with physical effects
(screening), we consider our approach to be
less {\em ad hoc} and more physically motivated
than some of the previous methods.

\begin{figure}[htb]
\begin{minipage}[b]{.49\linewidth}
 \centering\epsfig{file=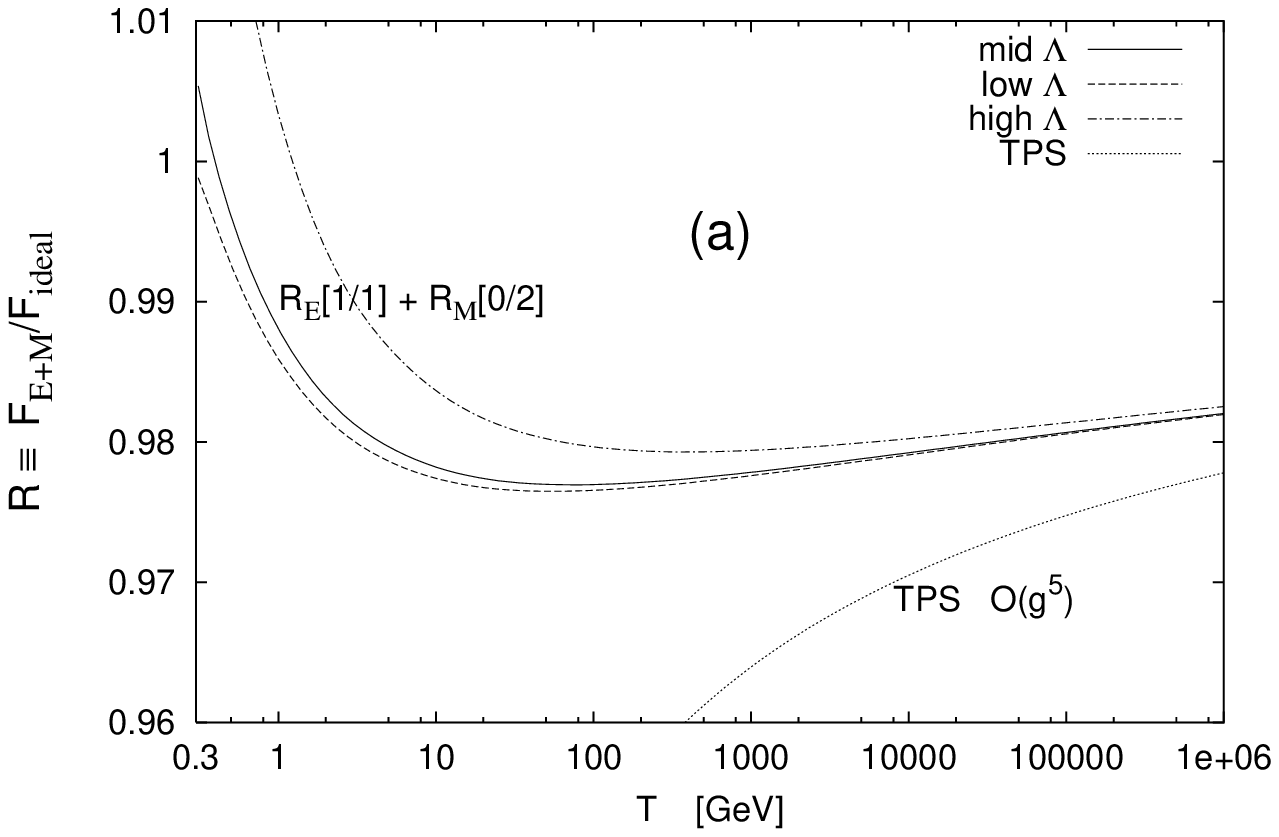,width=\linewidth}
\end{minipage}
\begin{minipage}[b]{.49\linewidth}
 \centering\epsfig{file=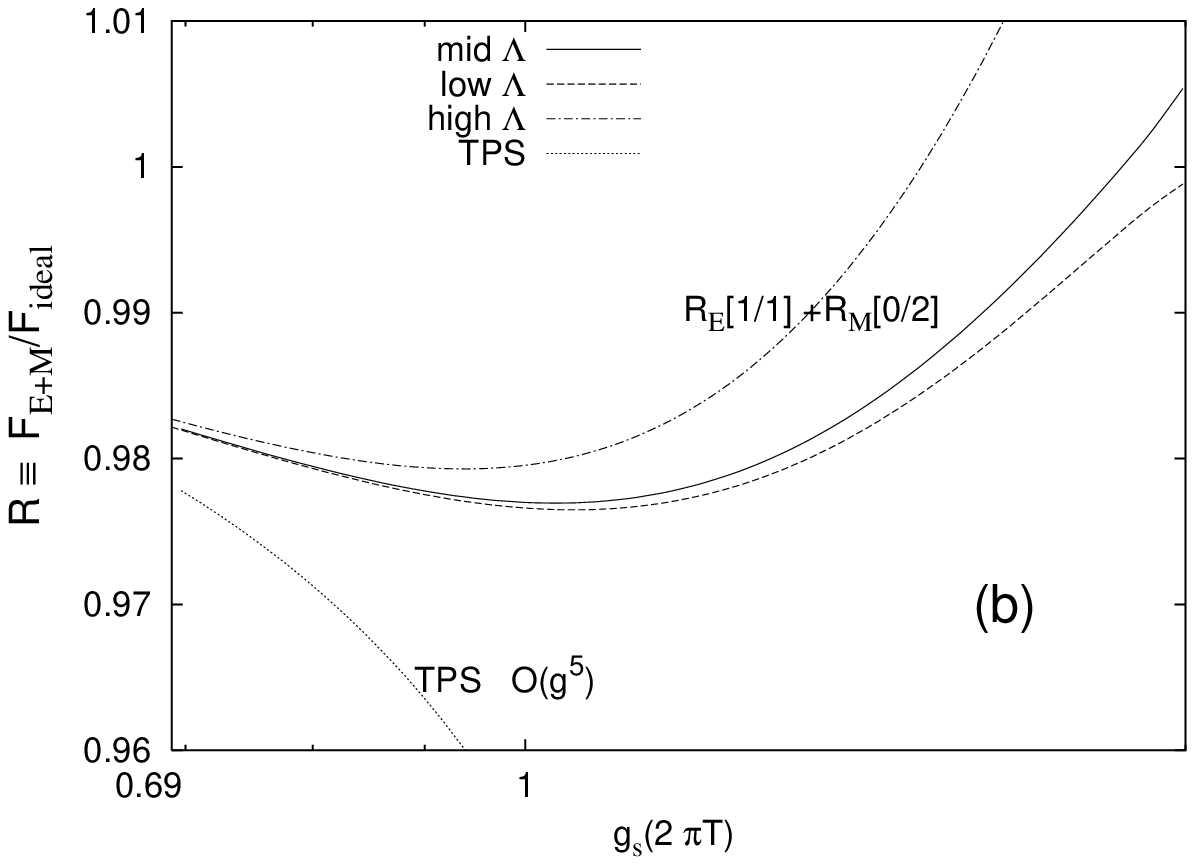,width=\linewidth}
\end{minipage}
\vspace{0.0cm}
\caption{\footnotesize The approximant $R_{E}[1/1]\!+\!R_{M}[0/2]$
as function of temperature (a) and of $g_s(2 \pi T)$ (b),
for three different choices 
of the relevant scales ($\mu_{\rm E}, \mu_{\rm M}\!
=\!\mu_m$, $\Lambda_{\rm E}$), as explained in the text.
For comparison, the TPS curve in powers of $g_s(2 \pi T)$
is included. In our case we have $g_s(2 \pi T) \approx 1.2 \ (1.0)$
for $T = 10 \ (100)$ GeV.}
\label{FEMvsTg}
\end{figure}

\begin{figure}[htb]
\begin{minipage}[b]{.49\linewidth}
 \centering\epsfig{file=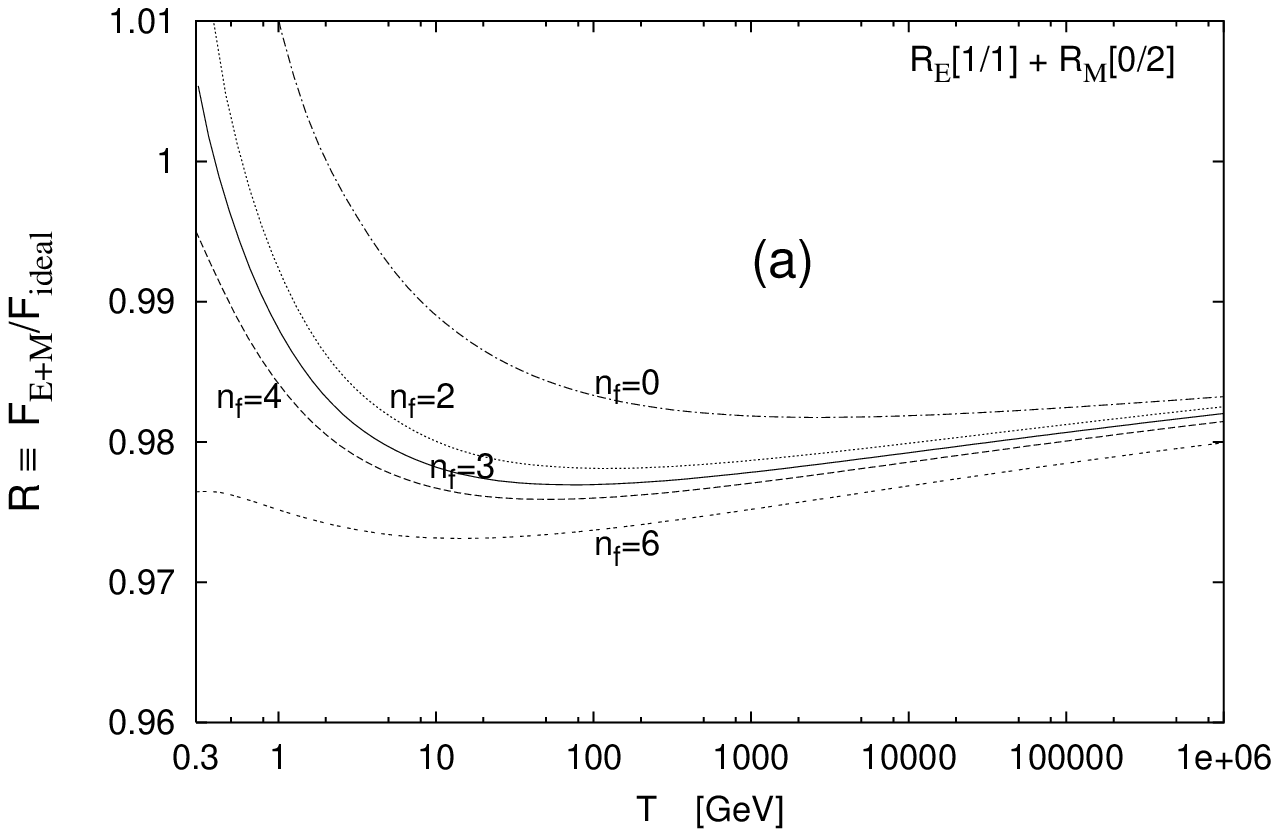,width=\linewidth}
\end{minipage}
\begin{minipage}[b]{.49\linewidth}
 \centering\epsfig{file=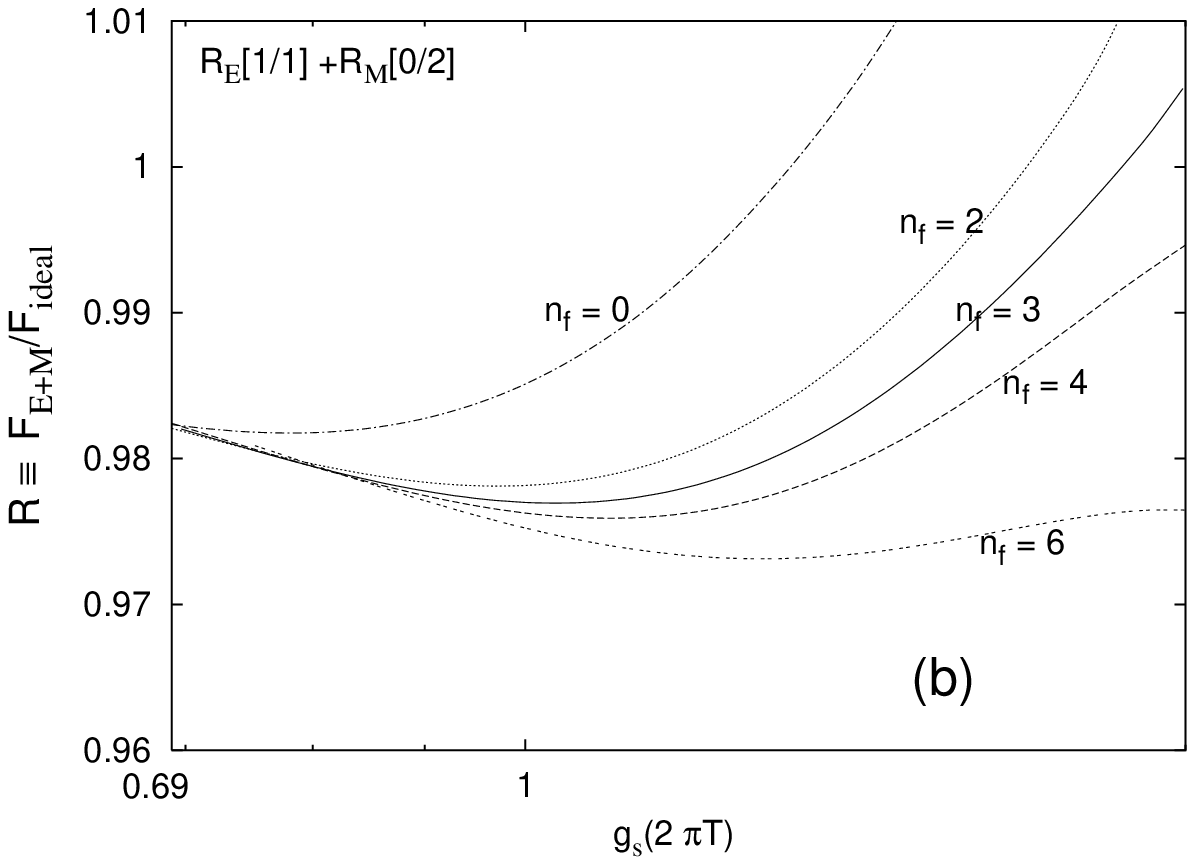,width=\linewidth}
\end{minipage}
\vspace{0.0cm}
\caption{\footnotesize Same as in Fig.~\ref{FEMvsTg}, but
for five different choices of the number of active massless quarks:
$n_f\!=\!0,2,3,4$, and $6$. The choice of the relevant scales is
mid $\Lambda$, as explained in the text. In all the previous
figures, we took the canonical value $n_f\!=\!3$.}
\label{FEMvsTgnf}
\end{figure}

\begin{figure}[htb]
\setlength{\unitlength}{1.cm}
\begin{center}
\epsfig{file=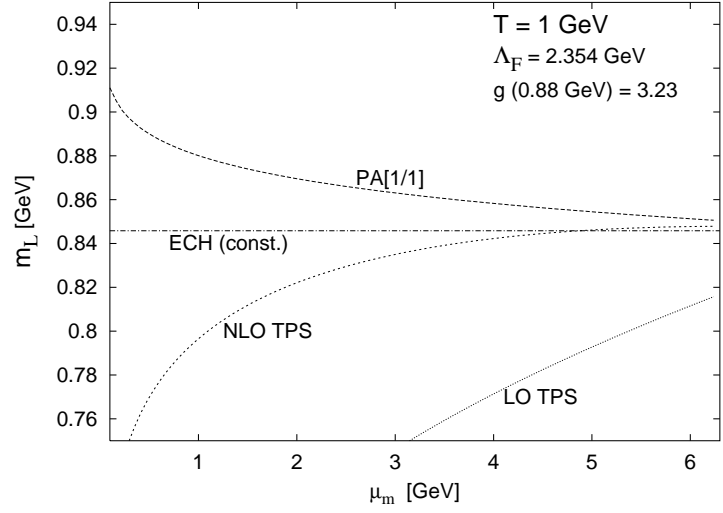, width=10.cm}
\end{center}
\vspace{0.0cm}
\caption{\footnotesize The screening mass $m_L$ in the
massless scalar $\phi^4$ theory as a function of the
RScl $\mu_m$, at $T\!=\!1$ GeV. Other details are
given in the text.}
\label{mLvsmuT=1}
\end{figure}

\begin{figure}[htb]
\begin{minipage}[b]{.49\linewidth}
 \centering\epsfig{file=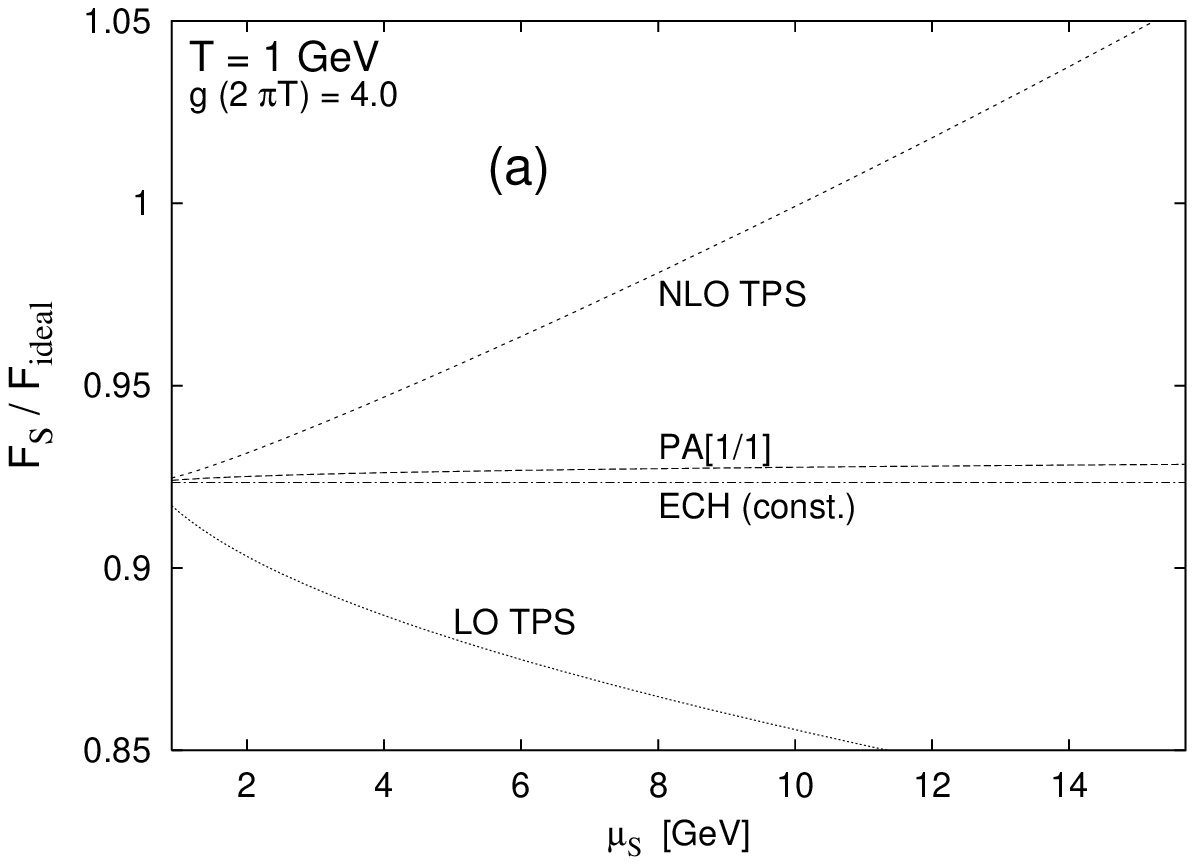,width=\linewidth}
\end{minipage}
\begin{minipage}[b]{.49\linewidth}
 \centering\epsfig{file=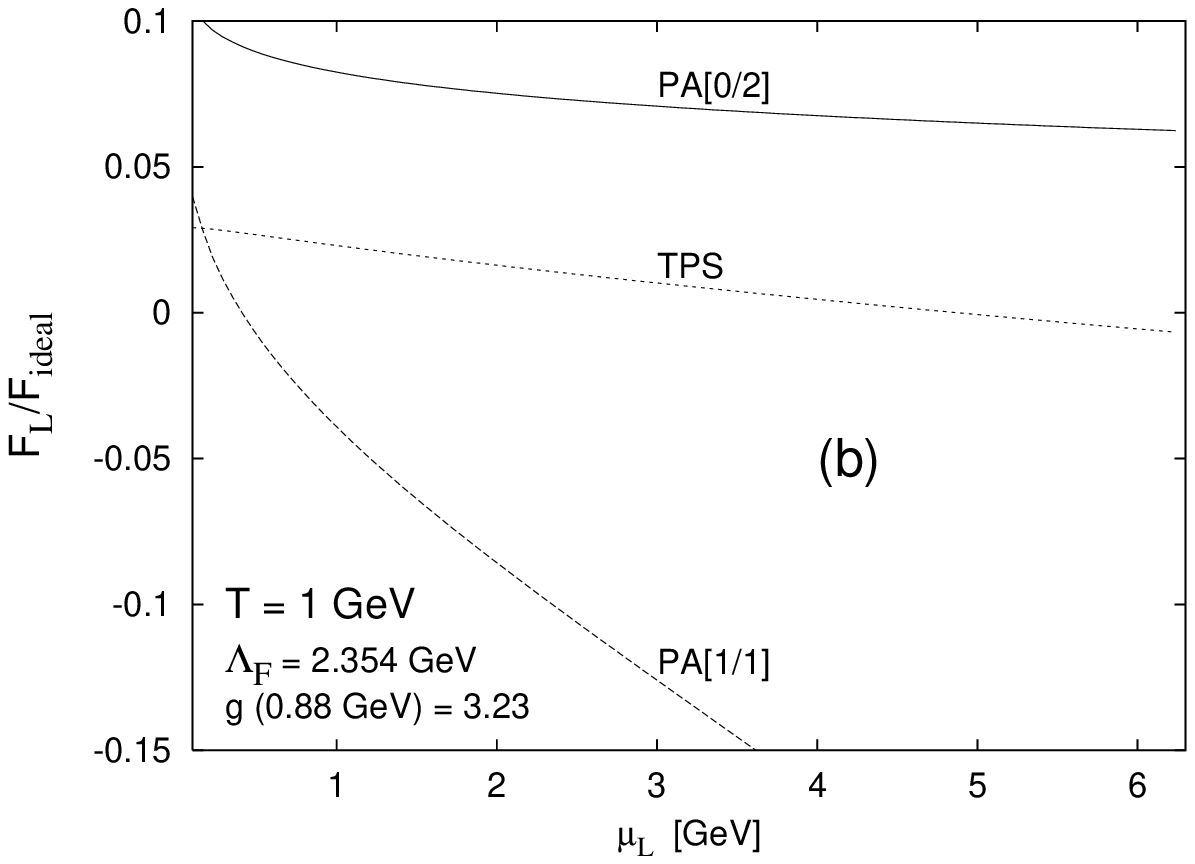,width=\linewidth}
\end{minipage}
\vspace{0.2cm}
\caption{\footnotesize 
The high energy $F_{\rm S}$ (a) and low energy $F_{\rm L}$ (b)
contributions to the free energy in $\phi^4$ theory
as functions of the corresponding RScl's.
Other details are given in the text.}
\label{F12vsmuT=1}
\end{figure}

\begin{figure}[htb]
\setlength{\unitlength}{1.cm}
\begin{center}
\epsfig{file=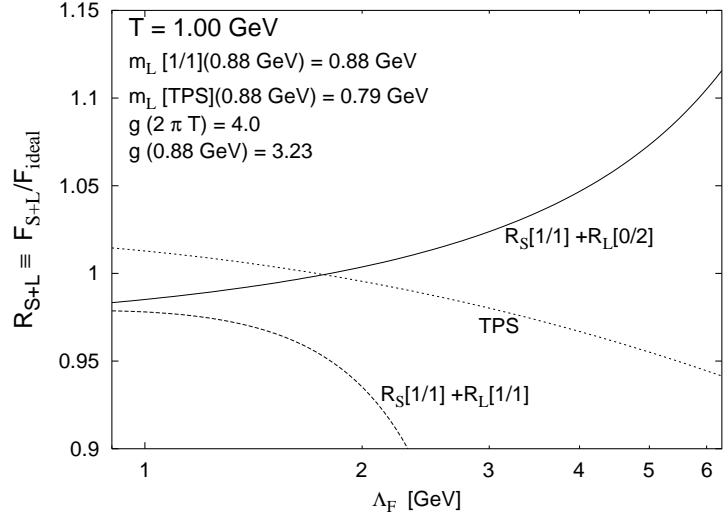, width=10.cm}
\end{center}
\vspace{0.0cm}
\caption{\footnotesize 
Various approximants for the normalized
sum $R_{\rm S+L}\!\equiv\!F_{\rm S+L}/F_{\rm ideal}$ as functions of the
factorization scale $\Lambda_F$. Other data
given in the figure and in the text.}
\label{F12vsLFT=1}
\end{figure}

\begin{figure}[htb]
\begin{minipage}[b]{.49\linewidth}
 \centering\epsfig{file=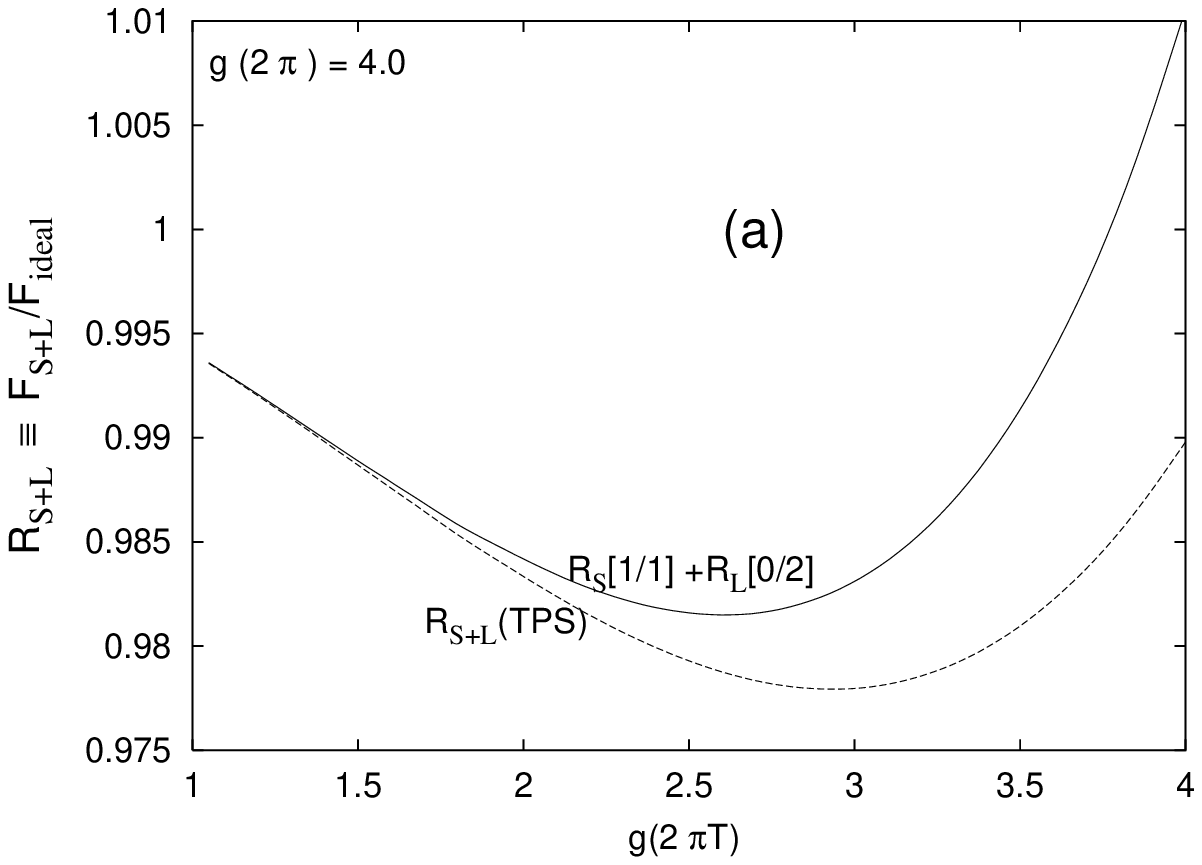,width=\linewidth}
\end{minipage}
\begin{minipage}[b]{.49\linewidth}
 \centering\epsfig{file=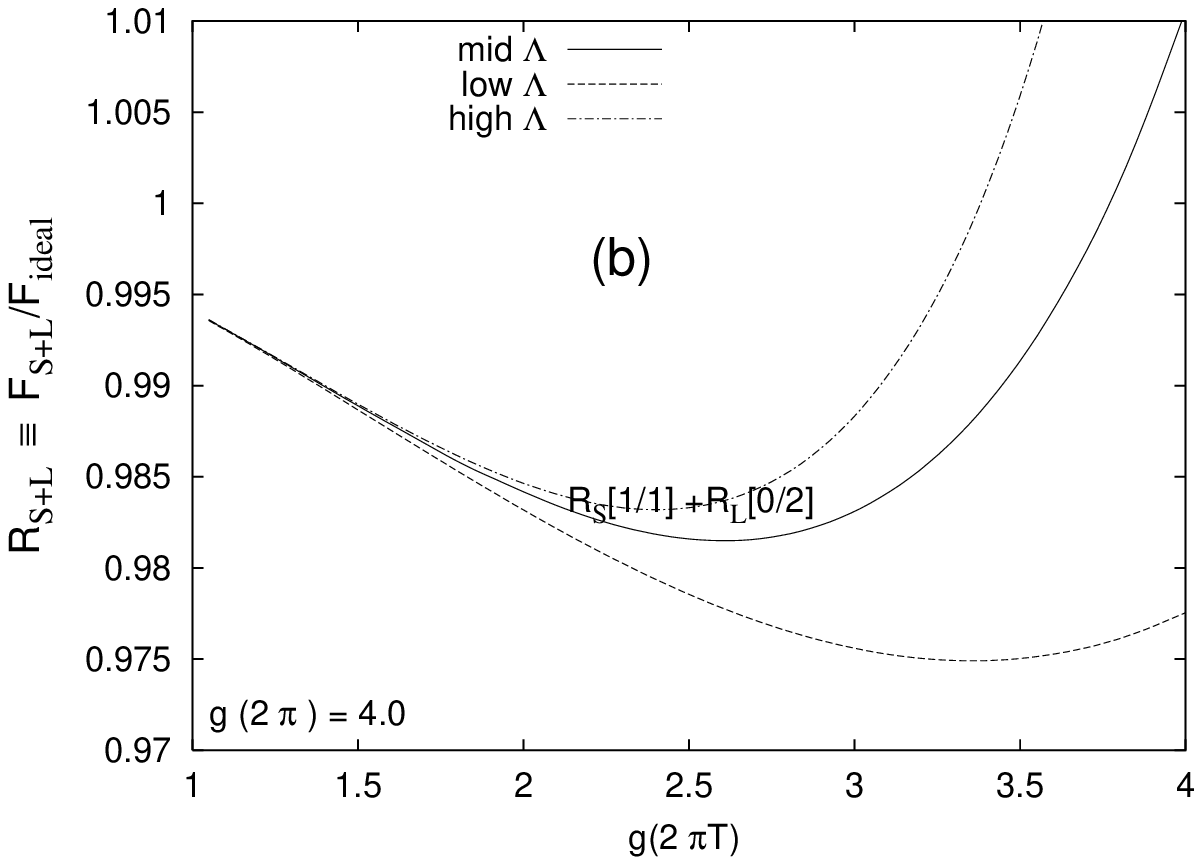,width=\linewidth}
\end{minipage}
\vspace{0.2cm}
\caption{\footnotesize 
The normalized sum $R_{\rm S+L}\!\equiv\!F_{\rm S+L}/F_{\rm ideal}$
as a function of $g(2 \pi T)$ (a).
(b) presents variation of the resummed result
$R_{\rm S}[1/1]+R_{\rm L}[0/2]$ for three choices of the relevant scales
($\mu_{\rm S}, \mu_{\rm L}\!=\!\mu_m, \Lambda_F$), 
as explained in the text.}
\label{F12vsg}
\end{figure}

As a consequence of the aforementioned
separation of the free energy into (two) parts, 
the underlying TPS's for the construction of PA's were 
of low order. Regarded from a purely numerical point of view,
this should alarm us and we should expect
significant (nonphysical) instabilities of the 
resulting low order PA's under variation of the RScl $\mu$,
and instabilites for the sum of the two parts
under variation of the factorization scale.
However, our results both in QCD and in $\phi^4$ theory
are remarkably stable under both variations. This 
confirms additionally that the described separation
forms a sound basis for the application of PA's.
Further, thus resummed values of $R = F/F_{\rm ideal}$
turn out to be below the value $1$ for a relatively
wide interval of the coupling parameter:
$g_s(2 \pi T) \leq 1.87$ 
[$\Leftrightarrow \ \alpha_s(2 \pi T) \leq 0.278$],
i.e., $T \geq 0.4$ GeV, in QCD with $n_f\!=\!3$;
see Fig.~\ref{FEMvsTg}(b). However, the method
probably breaks down already at lower 
$g_s(2 \pi T) \approx 1.0$-$1.2$,
i.e., $T \approx 10$-$100$ GeV, where
it gives the local minimum $R_{\rm min} \approx 0.977$
(if $n_f\!=\!3$).
The results at $T \agt 10$ GeV do not
contradict the results of lattice calculations.

\acknowledgments

The work of G.C. was supported by the FONDECYT (Chile)
Grants No.~1010094 and 7010094. R.K. would like to thank
the Department of Physics at the UTFSM, Valpara\'{\i}so,
for a very warm hospitality during the course of this work.
We are grateful to O.~Espinosa and D.~B\"odeker for
helpful discussions.




\newpage

\begin{thebibliography}{99}

\bibitem{Boyd:1995zg}
G.~Boyd, J.~Engels, F.~Karsch, E.~Laermann, C.~Legeland, 
M.~L\"utgemeier, and B.~Petersson,
Phys.\ Rev.\ Lett.\  {\bf 75}, 4169 (1995)
[arXiv:hep-lat/9506025];
Nucl.\ Phys.\ B {\bf 469}, 419 (1996)
[arXiv:hep-lat/9602007].

\bibitem{Engelsetal96}
J.~Engels, R.~Joswig, F.~Karsch, E.~Laermann, M.~L\"utgemeier, 
and B.~Petersson,
Phys.\ Lett.\ B {\bf 396}, 210 (1997)
[arXiv:hep-lat/9612018];
S.~Gottlieb {\it et al.},
Phys.\ Rev.\ D {\bf 55}, 6852 (1997)
[arXiv:hep-lat/9612020];
C.~W.~Bernard {\it et al.}  [MILC Collaboration],
Phys.\ Rev.\ D {\bf 55}, 6861 (1997)
[arXiv:hep-lat/9612025].


\bibitem{Karsch:2000ps}
F.~Karsch, E.~Laermann and A.~Peikert,
Phys.\ Lett.\ B {\bf 478}, 447 (2000)
[arXiv:hep-lat/0002003].

\bibitem{PeshierLevai}
A.~Peshier, B.~K\"ampfer, O.~P.~Pavlenko, and G.~Soff,
Phys.\ Rev.\ D {\bf 54}, 2399 (1996);
P.~Levai and U.~W.~Heinz,
Phys.\ Rev.\ C {\bf 57}, 1879 (1998)
[arXiv:hep-ph/9710463].


\bibitem{FrenkelParwani}
J.~Frenkel, A.~V.~Saa, and J.~C.~Taylor,
Phys.\ Rev.\ D {\bf 46}, 3670 (1992);
R.~Parwani and H.~Singh,
Phys.\ Rev.\ D {\bf 51}, 4518 (1995)
[arXiv:hep-th/9411065].


\bibitem{Braaten:1995cm}
E.~Braaten and A.~Nieto,
Phys.\ Rev.\ D {\bf 51}, 6990 (1995)
[arXiv:hep-ph/9501375].


\bibitem{ArnoldZhai}
P.~Arnold and C.~x.~Zhai,
Phys.\ Rev.\ D {\bf 50}, 7603 (1994)
[arXiv:hep-ph/9408276];
Phys.\ Rev.\ D {\bf 51} (1995) 1906
[arXiv:hep-ph/9410360];
C.~x.~Zhai and B.~Kastening,
Phys.\ Rev.\ D {\bf 52}, 7232 (1995)
[arXiv:hep-ph/9507380].

\bibitem{Braaten:1996ju}
E.~Braaten and A.~Nieto,
Phys.\ Rev.\ Lett.\  {\bf 76}, 1417 (1996)
[hep-ph/9508406];
E.~Braaten and A.~Nieto,
Phys.\ Rev.\ D {\bf 53} (1996) 3421
[hep-ph/9510408].


\bibitem{Andersen:1999sf}
J.~O.~Andersen, E.~Braaten, and M.~Strickland,
Phys.\ Rev.\ D {\bf 61}, 014017 (2000)
[arXiv:hep-ph/9905337];
J.~O.~Andersen, E.~Braaten, E.~Petitgirard and M.~Strickland,
arXiv:hep-ph/0205085.

\bibitem{Andersen:2000yj}
J.~O.~Andersen, E.~Braaten, and M.~Strickland,
Phys.\ Rev.\ D {\bf 63}, 105008 (2001)
[arXiv:hep-ph/0007159].


\bibitem{Kapusta}
J.~I.~Kapusta, Finite-temperature field theory
(Cambridge University Press, Cambridge, UK, 1989);
M.~Le Bellac, Thermal field theory
(Cambridge University Press, Cambridge, UK, 1996).

\bibitem{Karsch:1997gj}
F.~Karsch, A.~Patkos, and P.~Petreczky,
Phys.\ Lett.\ B {\bf 401}, 69 (1997)
[arXiv:hep-ph/9702376].

\bibitem{Chiku:1998kd}
S.~Chiku and T.~Hatsuda,
Phys.\ Rev.\ D {\bf 58}, 076001 (1998)
[arXiv:hep-ph/9803226].

\bibitem{BIR}
J.~P.~Blaizot, E.~Iancu and A.~Rebhan,
Phys.\ Lett.\ B {\bf 470}, 181 (1999)
[arXiv:hep-ph/9910309];
Phys.\ Rev.\ Lett.\  {\bf 83}, 2906 (1999)
[arXiv:hep-ph/9906340];
Phys.\ Rev.\ D {\bf 63}, 065003 (2001)
[arXiv:hep-ph/0005003];
Phys.\ Lett.\ B {\bf 523}, 143 (2001)
[arXiv:hep-ph/0110369].

\bibitem{Peshier:2000hx}
A.~Peshier,
Phys.\ Rev.\ D {\bf 63}, 105004 (2001)
[arXiv:hep-ph/0011250].

\bibitem{Drummond:1997cw}
I.~T.~Drummond, R.~R.~Horgan, P.~V.~Landshoff and A.~Rebhan,
Nucl.\ Phys.\ B {\bf 524}, 579 (1998)
[arXiv:hep-ph/9708426].

\bibitem{Drummond:1999si}
I.~T.~Drummond, R.~R.~Horgan, P.~V.~Landshoff and A.~Rebhan,
Phys.\ Lett.\ B {\bf 460}, 197 (1999)
[arXiv:hep-th/9905207].

\bibitem{Padebook}
George A. Baker, Jr. and Peter Graves-Morris,
{\it Pad\'e Approximants\/}, 2nd edition,
(Encyclopedia of Mathematics and Its Applications, Vol.~59),
edited by Gian-Carlo Rota
(Cambridge University Press, 1996).

\bibitem{Gardi:1996iq}
E.~Gardi,
Phys.\ Rev.\ D {\bf 56}, 68 (1997)
[arXiv:hep-ph/9611453].

\bibitem{CK2}
G.~Cveti\v c, Nucl. Phys. {\bf B517}, 506 (1998);
Phys. Rev. D {\bf 57}, R3209 (1998);
G.~Cveti\v c and R.~K\"ogerler, Nucl. Phys. {\bf B522}, 396 (1998);
G.~Cveti\v c, Nucl. Phys. B (Proc. Suppl.) {\bf 74}, 333 (1999);


\bibitem{CK1}
G.~Cveti\v c, Phys. Lett. B {\bf 486}, 100 (2000);
G.~Cveti\v c and R.~K\"ogerler,
Phys. Rev. {\bf D63} (2001) 056013.


\bibitem{Kastening:1997rg}
B.~Kastening,
Phys.\ Rev.\ D {\bf 56}, 8107 (1997)
[arXiv:hep-ph/9708219].


\bibitem{Hatsuda:1997wf}
T.~Hatsuda,
Phys.\ Rev.\ D {\bf 56}, 8111 (1997)
[arXiv:hep-ph/9708257].

\bibitem{lightbylight}
S.~J.~Brodsky, G.~P.~Lepage and P.~B.~Mackenzie,
Phys.\ Rev.\ D {\bf 28}, 228 (1983).
A.~L.~Kataev and V.~V.~Starshenko,
Phys.\ Rev.\ D {\bf 52}, 402 (1995)
[arXiv:hep-ph/9412305];
A.~L.~Kataev and V.~V.~Starshenko,
Mod.\ Phys.\ Lett.\ A {\bf 10}, 235 (1995)
[arXiv:hep-ph/9502348];
J.~R.~Ellis, I.~Jack, D.~R.~Jones, M.~Karliner and M.~A.~Samuel,
Phys.\ Rev.\ D {\bf 57}, 2665 (1998)
[arXiv:hep-ph/9710302].
F.~A.~Chishtie and V.~Elias,
Phys.\ Lett.\ B {\bf 521}, 434 (2001)
[arXiv:hep-ph/0107052];
C.~Contreras, G.~Cveti\v c, K.~S.~Jeong and T.~Lee,
arXiv:hep-ph/0203201.

\bibitem{Gross:1980br}
D.~J.~Gross, R.~D.~Pisarski and L.~G.~Yaffe,
Rev.\ Mod.\ Phys.\  {\bf 53}, 43 (1981).

\bibitem{Braaten:1994na}
E.~Braaten,
Phys.\ Rev.\ Lett.\  {\bf 74}, 2164 (1995)
[arXiv:hep-ph/9409434].


\bibitem{Appelquist}
T.~Appelquist and R.~D.~Pisarski,
Phys.\ Rev.\ D {\bf 23}, 2305 (1981).
N.~P.~Landsman,
Nucl.\ Phys.\ B {\bf 322}, 498 (1989).

\bibitem{Linde:px}
A.~D.~Linde,
Rept.\ Prog.\ Phys.\  {\bf 42}, 389 (1979);
A.~D.~Linde,
Phys.\ Lett.\ B {\bf 96}, 289 (1980).


\bibitem{ECH}
G.~Grunberg, Phys. Lett. {\bf 95B}, 70 (1980),
{\bf 110B}, 501(E) (1982);
{\bf 114B}, 271 (1982);
Phys. Rev. D {\bf 29}, 2315 (1984).

\bibitem{KKP}
A.~L.~Kataev, N.~V. Krasnikov, and A.~A. Pivovarov,
Nucl. Phys. {\bf B198}, 508 (1982).

\bibitem{Gupta}
A.~Dhar and V.~Gupta, Phys. Rev. D {\bf 29}, 2822 (1984);
V.~Gupta, D. V. Shirkov, and O. V. Tarasov, 
Int. J. Mod. Phys. A {\bf 6}, 3381 (1991).


\bibitem{Kleinert:rg}
H.~Kleinert, J.~Neu, V.~Schulte-Frohlinde, K.~G.~Chetyrkin and S.~A.~Larin,
Phys.\ Lett.\ B {\bf 272}, 39 (1991)
[Erratum-ibid.\ B {\bf 319}, 545 (1993)]
[arXiv:hep-th/9503230].

\end{thebibliography}
\end{document}